\documentclass{ieeeaccess}
\usepackage{cite}
\usepackage{amsmath,amssymb,amsfonts,amsthm}
\usepackage{algorithmic}
\usepackage{graphicx}
\usepackage{textcomp}
\usepackage{type1cm}
\usepackage{color}
\usepackage{bm,url}
\usepackage{caption,subcaption}
\usepackage{autobreak}
\newtheorem{theo}{Theorem}
\newtheorem{defi}{Definition}

\def\BibTeX{{\rm B\kern-.05em{\sc i\kern-.025em b}\kern-.08em
    T\kern-.1667em\lower.7ex\hbox{E}\kern-.125emX}}
\begin{document}
\history{Date of publication xxxx 00, 0000, date of current version xxxx 00, 0000.}
\doi{10.1109/ACCESS.2017.DOI}

\title{Key-Aggregate Searchable Encryption, Revisited: 
Formal Foundations for Cloud Applications, and Their Implementation}
\author{\uppercase{Masahiro Kamimura}\authorrefmark{1}, \IEEEmembership{Non-Member, IEEE},
\uppercase{Naoto Yanai}\authorrefmark{1}, \IEEEmembership{Member, IEEE}, 
\uppercase{Shingo Okamura}\authorrefmark{2}, \IEEEmembership{Member, IEEE}, 
\uppercase{Jason Paul Cruz}\authorrefmark{1} \IEEEmembership{Member, IEEE}} 
\address[1]{Graduate School of Information Science and Technology, Osaka University
1-5, Yamadaoka, Suita, Japan}
\address[2]{National Institute of Technology, Nara College, Japan}
\tfootnote{This paragraph of the first footnote will contain support 
information, including sponsor and financial support acknowledgment. For 
example, ``This work was supported in part by the U.S. Department of 
Commerce under Grant BS123456.''}

\tfootnote{This research was supported in part by the Japan Society for the Promotion of Science KAKENHI Number 18K18049, and Secom Science and Technology Foundation. We appreciate their support. We also would like to thank members of the study group ``Shin-Akarui-Angou-Benkyou-Kai" for the valuable discussions and helpful comments.}

\markboth
{Author \headeretal: Preparation of Papers for IEEE TRANSACTIONS and JOURNALS}
{Author \headeretal: Preparation of Papers for IEEE TRANSACTIONS and JOURNALS}

\corresp{Corresponding author: Naoto Yanai (e-mail: yanai@ist.osaka-u.ac.jp).}

\begin{abstract}
In the use of a cloud storage, sharing of data with efficient access control is an important requirement in addition to data security and privacy. Cui et al. (IEEE Trans. on Comp. 2016) proposed \textit{key-aggregate searchable encryption (KASE)}, which allows a data owner to issue an \textit{aggregate key} that enables a user to search in an authorized subset of encrypted files by generating an encrypted keyword called \textit{trapdoor}. While the idea of KASE is elegant, to the best of our knowledge, its security has never been discussed formally. 
%
In this paper, we discuss the security of KASE formally and propose provably secure schemes. The construction of a secure KASE scheme is non-trivial, and we will show that the KASE scheme of Cui et al. is insecure under our definitions. We first introduce our provably secure scheme, named \textit{first construction}, with respect to encrypted files and aggregate keys in a single-server setting. In comparison with the scheme of Cui et al., the first construction is secure without increased computational costs. Then, we introduce another provably secure scheme, named \textit{main construction}, with respect to trapdoors in a two-server setting. The main construction guarantees the privacy of a search, encrypted files, and aggregate keys. Considering 5,000 encrypted files, the first construction can finish search within three seconds and the main construction can finish search within six seconds.
\end{abstract}


\titlepgskip=-15pt

\maketitle

\section{Introduction} 
\label{Introduction}

\subsection{Background}

A cloud storage service provides a solution for storing, accessing, and sharing files over the Internet. 
However, such a service may be vulnerable and hence leak stored data without the permission or knowledge of the data owners. 
To prevent this leakage, data owners would want to encrypt their files before uploading them to a cloud storage. Searchable encryption~\cite{song2000practical} allows users to search over encrypted data with a chosen keyword without decryption of the encrypted data. Searchable encryption is suitable for a storage of data even in a vulnerable cloud storage service. In particular, even if control of a storage is exploited by an adversary, leakage of the stored data can be prevented by virtue of encrypting the data. To maximize the capabilities of the cloud
and features of cloud storage, data owners should also be able to
share their files to intended recipients. In recent years, several researchers ~\cite{popa2013multi,kiayias2016efficient,hamlin2018multi} created schemes that authorize other users to search over encrypted data, therefore sharing the data without decryption in the multi-user setting as well as encrypted search. 

On this background, this paper aims to introduce a scheme that authorizes for search in encrypted data, i.e., the searchability, \textit{efficiently}. More specifically, suppose that a data owner who owns a set of original data can issue a key that enables other users to search in a subset of its encrypted data without decryption. In such setting, the following features are desirable: (1) the data owner issues only a \textit{single short key} that is independent of both the number of encrypted data and the number of users who the encrypted data by the data owner, (2) and the encrypted data can be shared with users \textit{without changing or reproducing the encrypted data}. 
These features can make the operations of a cloud storage service efficient because the keys and encrypted data become easier to manage for both a data owner and users. We note that these \textit{features are not implied in previous systems,} namely, multi-user searchable encryption~\cite{bao2008private,hwang2007public,van2015multi,zhao2011multi} and multi-key searchable encryption~\cite{popa2013multi,hamlin2018multi}, which do not include efficient management described above in their scope. Therefore, achieving the features described above is a non-trivial problem.

\begin{figure}[!t]
  \begin{center}
    \includegraphics[clip,width=7.0cm]{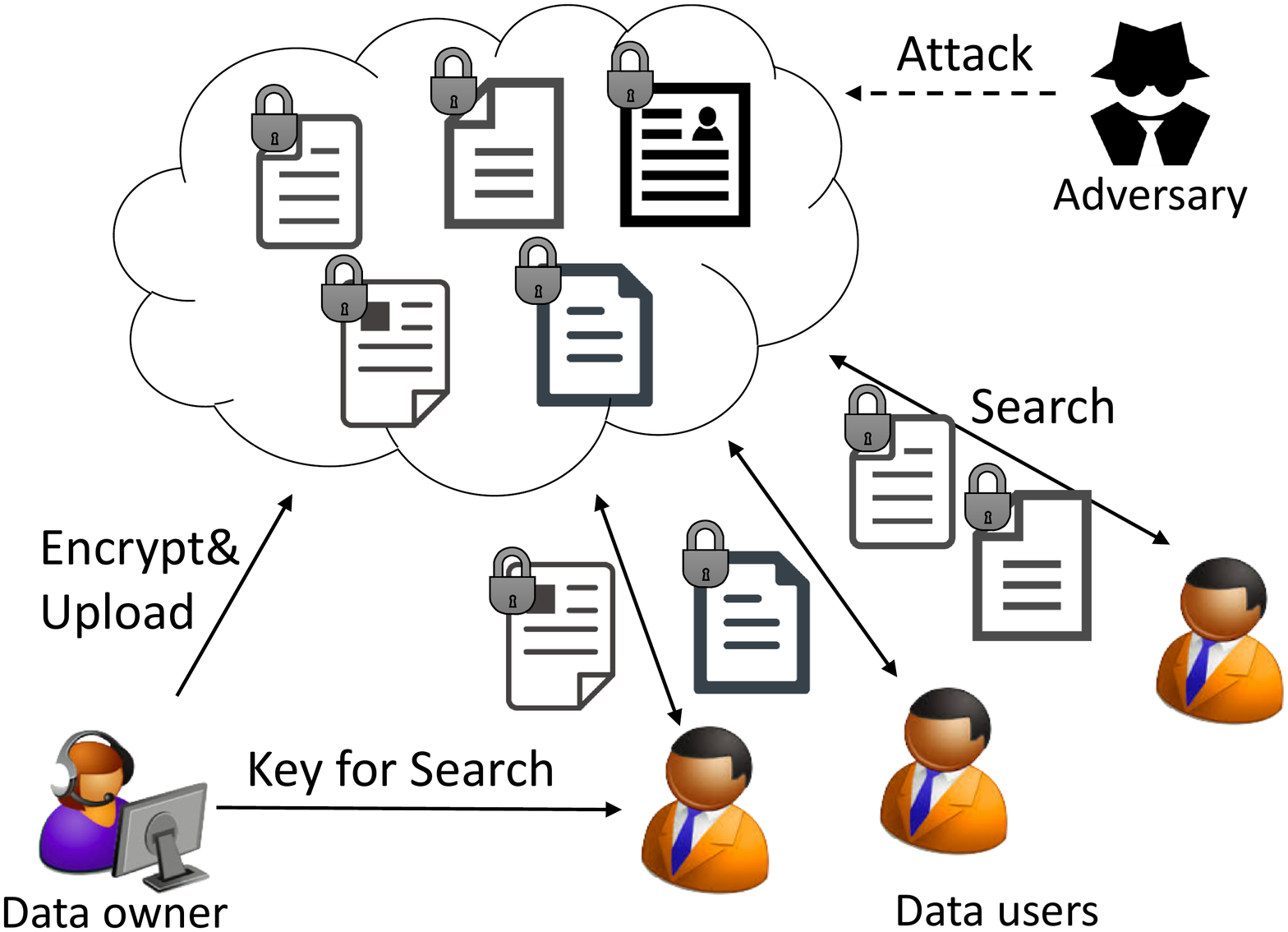}
    \caption{Overview of Key-Aggregate Searchable Encryption (KASE)}
    \label{fig:cloud-assisted_content-sharing_network}
  \end{center}
\end{figure}

Cui et al. ~\cite{cui2016key} proposed \textit{key-aggregate searchable encryption (KASE)} for the underlying purpose. 
An overview of KASE is shown in Figure~\ref{fig:cloud-assisted_content-sharing_network}. In KASE, a data owner issues aggregate keys that allow data users to search in authorized data only, i.e., data users generate trapdoors to search in encrypted data. 
The data sizes of ciphertexts and aggregate keys are independent of the number of data users the ciphertexts will be shared to, and the data size of ciphertexts is independent of the number of users. 
Therefore, KASE can improve the efficiency of the operations of a cloud storage service under the problem setting described above. 


However, Cui et al. did not provide formal definitions of the security of KASE and its security proofs. Moreover, Kiayias et al. ~\cite{kiayias2016efficient} introduced an attack against the scheme of Cui et al. in which encrypted keywords in ciphertexts are distinguishable for an adversary. To the best of our knowledge, no KASE scheme with formal security definitions and proofs has been introduced, even in subsequent works~\cite{li2018key,li2016verifiable}, making it an open problem. 

\subsection{Our contributions}

In this paper, we propose KASE schemes with provable security. To the best of our knowledge, these schemes are the first provably secure constructions. We also define a syntax and its security formally. We note that constructing a provably secure KASE scheme is \textit{non-trivial}. As will be described in Section~\ref{sec:difficulty} in details, KASE requires a data owner to control the searchability via only a single short key and without changing ciphertexts for each user. The algebraic structures that can be used to construct a KASE scheme are limited by the use of the single short key and keeping encrypted keywords.
we will newly show that the scheme of Cui et al. becomes insecure under our security definitions. 

In contrast, by shedding light on the mathematical features of the cryptographic primitives used to construct KASE, we construct a provably secure KASE scheme by tactically combining the existing instantiations of these cryptographic primitives. 
As will be described in Section~\ref{sec:constructions} in details, some instantiations of \textit{broadcast encryption}\cite{boneh2005collusion,fiat1993broadcast} and \textit{aggregate signatures}\cite{boneh2003aggregate} are generally combined as  basics. A simple combination called \textit{first construction} can provide \textit{keyword privacy} that guarantees the confidentiality of encrypted keywords and \textit{aggregate key unforgeability} that authorizes the searchability of keywords from a data owner to users. In addition to these two primitives, i.e., achieving the keyword privacy and the aggregate key unforgeability, the use of \textit{secret sharing}~\cite{shamir1979share} can provide \textit{trapdoor privacy} that guarantees the privacy of search for the users. We call the construction with secret sharing as \textit{main construction}.

Notably, computational cost and the size of ciphertexts in the first construction are identical to those of the scheme by Cui et al.~\cite{cui2016key}, but the first construction is provably secure. Considering 5,000 keywords, our two constructions can encrypt all keywords within one second, the first construction can perform search within three seconds, and the main construction can perform search within six seconds. We leave the construction of a generic scheme based on any instantiation of the primitives we used as an open problem.


\subsection{Potential Applications} 
KASE has many potential applications. We describe cloud-assisted content-sharing networks with cryptography~\cite{WWD13} and privacy-preserving authenticated communication in smart home environment~\cite{poh2019privhome} as example applications below. 

Content sharing networks are networks that are scalable in terms of the number of users, storage size, and network bandwidth. Cloud-assisted content sharing networks with cryptography mainly aim to enable both a service provider and users to control the privacy of data flexibly and scalably. According to Wu et al.~\cite{WWD13}, cloud-assisted content sharing networks with cryptography allows: 
(1) an individual to freely produce any number and any kind of online media, such as texts, images, and videos; (2) an individual to grant any access to his/her media to anyone at any time; (3) an individual to reveal a large number of attributes (e.g., age, address, and gender), some of which can be dynamic; and (4) an individual to share contents using various devices and bandwidths, and hence demand different access privileges for the same media. 
Wu et al. proposed an instanton of cloud-assisted content sharing networks with cryptography by utilizing attribute-based encryption~\cite{bethencourt2007ciphertext}. However, attribute-based encryption does not provide the searchability for ciphertexts, and hence the content of ciphertexts cannot be searched without decryption. This limitation make may lose the users' attention due to the lack of the searchability for encrypted contents. Moreover, attribute-based encryption requires a trusted third-party to generate secret keys, creating a potential single point of failure  in the entire system. 
Furthermore, the number of secret keys increases with the number of attributes, and thus the size of a storage and management cost of the keys also increases with the number of users and the management of authorization for content sharing. 
In contrast, KASE provides searchability of contents of ciphertexts and a fixed size of keys for both a data owner and users, solving the problems of cloud-assisted content-sharing networks described above. Therefore, KASE is desirable for use in cloud-assisted content-sharing networks. 


The privacy-preserving authenticated communication in smart home environment by Poh et al.~\cite{poh2019privhome} is an application where a user can securely utilize smart devices that accumulate private information, such as sleeping patterns and medical information.
Poh et al. realized such an application in the single-user setting by integrating searchable encryption and authenticated key-establishment protocol. 
However, smart devices are continuously becoming more popular, and thus the use of smart devices in the multi-user setting should be considered.
For example, in a case with a large number of users, e.g., employees in a workplace, each employee has a separated data access per device. KASE can be used to control access of each employee to devices and their data by the use of a single key. Thus, a more efficient and attractive usage of device management can be expected by virtue of KASE.


\subsection{Related works}

Following the original KASE proposed by Cui et al.~\cite{cui2016key}, many KASE schemes have been proposed in \cite{li2018key,li2016verifiable,mukti2018mulkase}. 
We note that the both schemes are essentially the same as the construction by Cui et al. and their security have never been proven. Liu et al. \cite{li2016verifiable} introduced the verifiability of search results, and Li et al. \cite{li2018key} discussed situations with multiple data owners. 
Meanwhile, Mukti et al.~\cite{mukti2018mulkase} proposed KASE with multiple data owners and discussed its security formally. Unfortunately, their definitions are incomplete and their proofs are wrong. In particular, in their syntax, any keyword can be searched as long as the keyword is included as a part of ciphertexts, even if its searched documents are the outside of his/her authorized documents. Moreover, in their proposed construction, the bilinear maps are nested for the test algorithm, which is an unworkable setting for bilinear maps. 


As a special research alleviating the conditions of KASE, Zhou et al. \cite{zhou2018file} proposed a searchable and secure scheme in the situation where remote sensor devices encrypt data. This scheme individually changes the key for aggregate key issuance and the key for data encryption.
However, this feature raises an issue of increased key management for a data owner.
Patranabis et al. \cite{Patranabis2017key} also considered variants of KASE, but their scheme has no search delegation to other users.
Therefore, the situation that KASE normally handles is different, so it can be said that it is different from the problem dealt with in this paper.

In a framework of conventional searchable encryption, the multi-user setting~\cite{bao2008private,hwang2007public,van2015multi,zhao2011multi,zhang2018efficient,wu2018verifiable} and the multi-key setting~\cite{popa2013multi,hamlin2018multi,wang2017idcrypt} have been known to control the searchability for each document. However, an efficient control of the searchability, such as issuance of aggregate keys which is one of the main problems in KASE, is out of the scope of such settings. 


Searchable attribute-based encryption (SABE)~\cite{liang2015searchable,zheng2014vabks,yin2019cp,alrawais2017attribute,wang2018efficiently} is an encryption scheme similar to KASE. SABE is a searchable encryption scheme in which documents corresponding to attributes are searchable for users who own secret keys of the attributes. While SABE provides searchability along with attributes for each document, the size of secret keys depends on the number of attributes, i.e., the key size is linear. This problem in SABE is a different problem from KASE. 

As additional related works, key-aggregate cryptosystems~\cite{guo2018key,chu2014kac,patranabis2017kac} outsource the decryption of data. Among them, Patranabis et al.~\cite{patranabis2017kac} showed a provably secure scheme. However, the searchability for ciphertexts is out of the scope in these works. 
Since the searchability is not implied generally, constructing a provably secure KASE remains an open problem.

\subsubsection{Paper Organization}

The rest of this paper is organized as follows. 
The mathematical background to understanding this work is described in Section~\ref{Preliminaries}, and then a definition of KASE and the technical difficulty in constructing a KASE scheme is presented in Section~\ref{Key Aggregate Searchable Encryption}. 
The main idea to overcome the difficulty and the proposed KASE schemes are discussed in Section~\ref{sec:constructions}, and then their security proofs are presented in Section~\ref{Proofs}. 
Implementation, evaluation, and analysis are discussed in Section~\ref{Discussion}. 
Finally, the conclusion and future direction are presented in Section~\ref{Conclusion}.

\section{Preliminaries} \label{Preliminaries}

In this section, we present the background to groups with bilinear maps and their security assumptions. 

\subsection{Bilinear maps}
\label{bilinear_maps}

The proposed schemes are based on bilinear maps and bilinear groups. 
In this paper, we recall the standard description defined in~\cite{joux2000one}. 
Here, let $\mathbb{G}$, $\mathbb{H}$, and $\mathbb{G}_T$ be groups with the same prime order $p$. 
We then define bilinear groups $(\mathbb{G}, \mathbb{H}, \mathbb{G}_T)$ and bilinear maps $e:\mathbb{G}\times \mathbb{H}\to \mathbb{G}_T$ as follows: 
\begin{enumerate}
\item Bilinear groups $(\mathbb{G}, \mathbb{H}, \mathbb{G}_T)$ are two cyclic groups with a prime order $p$.
\item $g\in \mathbb{G}$ and $h \in \mathbb{H}$ are generators of $\mathbb{G}$ and $\mathbb{H}$, respectively. 
\item A bilinear map $e:\mathbb{G}\times \mathbb{H} \to \mathbb{G}_T$ is a map with the following properties:

\begin{enumerate}
\item Bilinearity. For any value $u\in \mathbb{G}$, $v \in \mathbb{H}$ and $a,b\in \mathbb{Z}_p^*$, $e(u^a,v^b)=e(u,v)^{ab}$ holds.
\item Non-degeneracy. $e(g,h)\not=1$ holds, where $1$ means an identity element over $\mathbb{G}_T$.
\item Computability. For any value $u\in \mathbb{G}$ and $h \in \mathbb{H}$, $e(u,v)$ can be calculated efficiently.
\end{enumerate}

\end{enumerate}
When $\mathbb{G} = \mathbb{H}$, bilinear groups are said to be \textit{symmetric} and denoted by $(\mathbb{G}, \mathbb{G}_T)$ for the sake of convenience. 
Likewise, $\mathbb{G} \neq \mathbb{H}$, bilinear groups are said to be \textit{asymmetric} and denoted by $(\mathbb{G}, \mathbb{H}, \mathbb{G}_T)$. 

\subsection{Complexity Assumptions}
\label{complexity_assumptoins}
In this section, we define security assumptions utilized in the proposed schemes.

\subsubsection{BDHE Assumption in $(\mathbb{G}, \mathbb{G}_T)$}
The bilinear Diffie-Hellman exponentiation (BDHE) assumption in $(\mathbb{G}, \mathbb{G}_T)$ is an assumption introduced by Boneh et al.~\cite{boneh2005collusion}.

\begin{defi}[$(\epsilon, l)$-BDHE Assumption in $(\mathbb{G}, \mathbb{G}_T)$]
We say the $l$-BDHE problem in $(\mathbb{G}, \mathbb{G}_T)$ with a security parameter $1^k$ as, 
for a given $(g,h,g^\alpha,g^{\alpha^2},...,g^{\alpha^l},g^{\alpha^{l+2}},...,g^{\alpha^{2l}}, Z)$ with uniformly random $(g,h)\in \mathbb{G}$, $\alpha\in \mathbb{Z}_p^*$ and $(\mathbb{G}, \mathbb{G}_T)$ as input, 
determining whether $Z\in \mathbb{G}_T$ is $e(g^{\alpha^{l+1}},h)$ or a random value $R$. 
We say that a polynomial time algorithm $\mathcal{A}$ can solve the $l$-BDHE problem in $(\mathbb{G}, \mathbb{G}_T)$ with an advantage $\epsilon$ if the following relation holds:
\begin{eqnarray}
|Pr[\mathcal{A}(g,h,\bm{y}_{g,\alpha,l},e(g^{\alpha^{l+1}},h), \mathbb{G}, \mathbb{G}_T)=0]\nonumber \\
-Pr[\mathcal{A}(g,h,\bm{y}_{g,\alpha,l},R, \mathbb{G}, \mathbb{G}_T)=0]|\ge \epsilon,  \nonumber
\end{eqnarray}
where $\bm{y}_{g,\alpha,l}=(g^\alpha,g^{\alpha^2},...,g^{\alpha^l},g^{\alpha^{l+2}},...,g^{\alpha^{2l}})$.
We say the $(\epsilon, l)$-BDHE assumption holds in $(\mathbb{G}, \mathbb{G}_T)$ if there is no polynomial-time algorithm that can solve the $l$-BDHE problem in $(\mathbb{G}, \mathbb{G}_T)$ with $\epsilon$.
\end{defi}

\subsubsection{DHE Assumption in $\mathbb{G}$}
The Diffie-Hellman exponentiation (DHE) assumption in $\mathbb{G}$ is an assumption introduced by Herranz et al.~\cite{herranz2012short}.

\begin{defi}[$(\epsilon, l)$-DHE Assumption in $\mathbb{G}$]
We say the $l$-DHE problem with a security parameter $1^k$ as, 
for a given $(g,g^\alpha,g^{\alpha^2},...,g^{\alpha^l},g^{\alpha^{l+2}},...,g^{\alpha^{2l}})$ with uniformly random $g\in \mathbb{G}$, $\alpha\in \mathbb{Z}_p^*$ and $(\mathbb{G}, \mathbb{G}_T)$ as input, 
computing $g^{\alpha^{l+1}}$. 
We say that a polynomial time algorithm $\mathcal{A}$ can solve the $l$-DHE problem in $\mathbb{G}$ with an advantage $\epsilon$ if the following relation holds: 
\[
Pr[\mathcal{A}(g,\bm{y}_{g,\alpha,l},g^{\alpha^{l+1}}, \mathbb{G}, \mathbb{G}_T)]\ge \epsilon, 
\]
where $\bm{y}_{g,\alpha,l}=(g^\alpha,g^{\alpha^2},...,g^{\alpha^l},g^{\alpha^{l+2}},...,g^{\alpha^{2l}})$.
We say the $(\epsilon, l)$-DHE assumption holds in $\mathbb{G}$ if there is no polynomial-time algorithm that can solve the $l$-DHE problem in $\mathbb{G}$ with $\epsilon$. 
\end{defi}

\subsubsection{XDH Assumption in $(\mathbb{G},\mathbb{H})$}
The external Diffie-Hellman (XDH) assumption in $(\mathbb{G},\mathbb{H})$ is an assumption introduced in~\cite{ateniese2005untraceable,ballard2005correlation}. 
Note that, unlike the BDHE assumption and the DHE assumption described above, the XDH assumption holds on only \textit{asymmetric} bilinear groups $(\mathbb{G}, \mathbb{H}, \mathbb{G}_T)$. 

\begin{defi}[$\epsilon$-XDH Assumption in $(\mathbb{G},\mathbb{H})$] 
We say the XDH problem in $(\mathbb{G},\mathbb{H})$ as, 
for a given $(g,h,g^a,g^b,Z)$ with uniformly random $g\in \mathbb{G}, h\in \mathbb{H}$, $(a,b)\in \mathbb{Z}_p^*$ and $(\mathbb{G}, \mathbb{H}, \mathbb{G}_T)$ as input, 
determining whether $Z\in \mathbb{G}$ is $g^{ab}$ or is a random value $R$. 
We say that a polynomial time algorithm $\mathcal{A}$ can solve XDH problem in $(\mathbb{G},\mathbb{H})$ with advantage $\epsilon$ if the following relation holds:
\begin{eqnarray}
|Pr[\mathcal{A}(g,h,g^a,g^b,g^{ab}, \mathbb{G}, \mathbb{H}, \mathbb{G}_T)=0]\nonumber \\
-Pr[\mathcal{A}(g,h,g^a,g^b,R, \mathbb{G}, \mathbb{H}, \mathbb{G}_T)=0]|\ge \epsilon. \nonumber
\end{eqnarray}
We say $\epsilon$-XDH assumption holds in $(\mathbb{G},\mathbb{H})$ if there is no polynomial-time algorithm that can solve the XDH problem in $(\mathbb{G},\mathbb{H})$ with $\epsilon$. 
\end{defi}

\section{Key-Aggregate Searchable Encryption(KASE)} \label{Key Aggregate Searchable Encryption}

In this section, we describe the main problem statement of key-aggregate searchable encryption (KASE)~\cite{cui2016key}. 
Then, we newly define a syntax of algorithms and the security for KASE. 
These definitions are our contributions. 

\subsection{Problem Statement}
In key-aggregate searchable encryption (KASE)~\cite{cui2016key}, 
a data owner provides a ``single-and-short'' aggregate key that enables a user to access documents for authorization of search. 
Each user, which we call data user for the sake of convenience, is given an aggregate key as secret information and then generates a ``single-and-short'' trapdoor to search for a keyword from the documents.
In doing so, the following requirements should be satisfied by KASE: 

\begin{itemize}
\item Searchability:
A user can generate trapdoors for any keyword to search in encrypted documents. 

\item Compactness:
The size of both aggregate keys and trapdoors should be independent of the number of documents and number of users. 
In addition, the size of encrypted keywords should be independent of the number of users. 

\item Keyword Privacy:
An adversary cannot extract information about the original keywords from encrypted documents. 
That is, a person who does not have an aggregate key corresponding to indexes of the documents cannot get any information from the encrypted keyword.

\item Aggregate Key Unforgeability:
An adversary cannot search for any keyword without authorization from a data owner. 
That is, an adversary cannot perform keyword search over the documents that are not related to the known aggregate key and it cannot generate new aggregate keys for other sets of documents from the known keys.
\end{itemize}
These requirements are also shown in the original work of KASE~\cite{cui2016key}. 
The compactness and the searchability are functionality for KASE while the keyword privacy and the aggregate key unforgeability are security for KASE.

As described above, the compactness must be satisfied because the main motivation of KASE is to keep aggregate keys and trapdoors short while providing the searchability.
We note that \textit{the size of encrypted keywords may depend on the number of documents} because the data size of the encrypted keywords with respect to the number of users is the out of the scope of efficient.
The keyword privacy is a requirement that prevents an adversary from getting information contained in encrypted documents. 
Meanwhile, aggregate key is a new notion required in KASE and has not been discussed in general searchable encryption. 
However, because documents outside the scope of authorization should be unsearchable, the aggregate key unforgeability should be discussed. Even if the keyword privacy is satisfied, there is a possibility that an adversary can search documents outside the scope of authorization. 
Thus, both the keyword privacy and the aggregate key unforgeability should be discussed.

As another security requirement, the following should be considered: 
\begin{itemize}
\item Trapdoor Privacy:
An adversary cannot determine a keyword embedded in the given trapdoor. 
That is, even when a user asks an untrusted cloud server to search, 
the server cannot obtain the keyword except for the search result.
\end{itemize}
We note that the trapdoor privacy is an additional security requirement, 
i.e., only a few schemes~\cite{cao2014privacy,kiayias2016efficient,arriaga2014trapdoor} satisfy the trapdoor privacy even in the conventional searchable encryption. 
However, satisfying the trapdoor privacy is an important feature. When keywords embedded in trapdoors are revealed, the original keywords can be analyzed from encrypted documents by looking up the search results. 
In other words, if the trapdoor privacy is unsatisfied, then the keyword privacy may be threatened.
Thus, we consider that a KASE scheme should satisfy all requirements described above.

To the best of our knowledge, no provably secure KASE scheme or formal security definitions have been proposed. 
Thus, in this paper, we define the requirements above formally. 


\subsection{Algorithms}
\label{algorithms}

The algorithms of KASE are defined as follows: 

\begin{itemize}
\item $params\gets Setup(1^\lambda,n)$:
This algorithm is run by a cloud service provider to set up the scheme. 
On input of a security parameter $1^\lambda$ and the maximum number $n$ of possible documents owned by a data owner, 
the algorithm outputs a public parameter $params$. 

\item $sk\gets KeyGen(params)$:
This algorithm is run by a data owner to generate a secret key $sk$.

\item $c_{i,l}\gets Encrypt(params, sk, i, w_l)$: 
This algorithm is run by a data owner to encrypt a keyword which belongs to the $i$th document and generate an encrypted keyword.  
On input of the data owner's secret key $sk$, an document index $i$ and a keyword $w_l\in \mathcal{KS}$ 
where $\mathcal{KS}$ is a keyword space, 
the algorithm outputs an encrypted keyword $c_{i,l}$. 

\item $k_{agg}\gets Extract(params, sk,S)$: 
This algorithm is run by a data owner to generate an aggregate key for delegating the keyword search capability for
a certain set of documents to other data users. 
On input of the data owner's secret key $sk$ and a set $S$ of indexes of documents, 
the algorithm outputs an aggregate key $k_{agg}$.

\item $Tr\gets Trapdoor(params, k_{agg},S,w_l)$:
This algorithm is run by a data user who performs the keyword search. 
On input of an aggregate key $k_{agg}$ and a keyword $w_l$, the algorithm outputs a single trapdoor $Tr$. 

\item $Tr_i\gets Adjust(params, i,S,Tr,\{f_{1,i}\}_{i\in [1,m_1]})$: 
This algorithm is run by a cloud server to adjust the given aggregate trapdoor for each document. 
On input of a set $S$ of indexes of documents, the index $i$ of the target document, an aggregate trapdoor $Tr$ and auxiliary functions $\{f_{i}\}_{i\in [1,m_1]}(m_1 \in \mathbb{N})$ possibly,
the algorithm outputs each trapdoor $Tr_i$ for the $i$th target document in $S$.

\item $b\gets Test(params, Tr_i,S,c_{i,l},\{f_{2,i}\}_{i\in [1,m_2]})$: 
This algorithm is run by a cloud server to perform keyword search over an encrypted document. 
On input of a trapdoor $Tr_i$, the document index $i$ and auxiliary functions $\{f_{i}\}_{i\in [1,m_2]}(m_2 \in \mathbb{N})$ possibly, 
the algorithm outputs $true$ or $false$ to denote whether the $i$th document contains 
the keyword $w_l$. 
\end{itemize}

We note that the syntax above represents a multi-server setting that includes multiple cloud servers. The syntax can contain multiple servers by setting auxiliary functions separately for each cloud server in the Adjust and Test algorithms.

We define the correctness of the syntax of KASE as follows: 
\begin{defi}[Correctness]
For any document
containing the keyword $w_l$ with index $i \in S$, 
We say that a KASE scheme satisfies the correctness if the following statement holds: 
for all $1^\lambda, n\in \mathbb{N}, i \in [1,n], w_l\in \mathcal{KS}$, when a public parameter $params\gets Setup(1^\lambda,n)$ and a secret key $sk\gets KeyGen(params), c_{i,l}\gets Encrypt(params, sk,i,w_l)$ are used, $Test(params, Tr_i, S,c_{i,l},\{f_{2,i}\}_{i\in [1,m_2]})= true$ if $Tr \gets Trapdoor(params, k_{agg},S,w_l)$ and $Tr_i\gets  Adjust(params, i, S, Tr,\{f_{1,i}\}_{i\in [1,m_1]})$. 
\end{defi}

The correctness defined above imposes the searchability of KASE because the correctness guarantees that a data user can search for any keyword without decryption. Moreover, the syntax described above is identical to the abstraction of the algorithms proposed in the original work of KASE~\cite{cui2016key} except that the symmetric key setting instead of the public key setting in~\cite{cui2016key}.

We define the compactness of KASE as follows: 
\begin{defi}[Compactness]
We say that KASE satisfies the compactness if the sizes of both aggregate keys and trapdoors are independent of the number $n$ of encrypted documents and the number $m$ of data users, i.e., $\mathcal{O}(1)$ with respect to $n$ and $m$, and the size of encrypted keywords is independent of $m$, i.e., $\mathcal{O}(n)$ with respect to $n$ and $m$. 
\end{defi}

\subsection{Security Definitions}
\label{security_definitions}

In this section, we define three security requirements for KASE, namely, the keyword privacy, the aggregate key unforgeability, and the trapdoor privacy. 
The security requirements are defined by the following game between an adversary $\mathcal{A}$ and a 
challenger $\mathcal{C}$. 
For each game, 
both $\mathcal{C}$ and $\mathcal{A}$ are given $(1^\lambda,n)$ as input, 
$\mathcal{A}$ is allowed to get aggregate keys, encrypted keywords, and trapdoors in the query phase by accessing the key extraction oracle $\mathcal{O}_{Extract}$, the encryption oracle $\mathcal{O}_{Encrypt}$, and the trapdoor oracle $\mathcal{O}_{Trapdoor}$, respectively. 
In particular, $\mathcal{A}$ accesses each oracle as follows: 
\begin{itemize}
\item $\mathcal{O}_{Extract}$: by taking $S\subseteq [1,n]$ as input, return $k_{agg}\gets Extract(params,sk,S)$. 
\item $\mathcal{O}_{Encrypt}$: by taking $i\in [1,n],w_l\in \mathcal{KS}$ as input, return $c_{i,l}\gets Encrypt(params,sk,i,w_l)$. 
\item $\mathcal{O}_{Trapdoor}$: by taking $S\subseteq [1,n], w_l \in \mathcal{KS}$ as input, return $Tr\gets Trapdoor(params,Extract(params,sk,S),\\S,w_l)$. 
\end{itemize}

\begin{defi}[$(\epsilon,n)$-Keyword Privacy]
In this game, an adversary $\mathcal{A}$ tries to distinguish a challenge keyword or a random keyword from a challenge encrypted keyword.
\begin{itemize}
\item Init:
$\mathcal{A}$ declares the index $i^* \in [1,n]$ of a challenge document used in the guess phase and sends it to $\mathcal{C}$. 

\item Setup:
$\mathcal{C}$ generates $params\gets Setup(1^\lambda,n)$ and $sk\gets KeyGen (params)$, and sends $params$ to $\mathcal{A}$. 

\item Query:
$\mathcal{A}$ can query to $\mathcal{O}_{Extract}$ at most $n-1$ times\footnote{If $\mathcal{A}$ can access to $\mathcal{O}_{Extract}$ more than $n$ times, $\mathcal{A}$ can generate trapdoors for every index. $\mathcal{A}$ can trivially break any scheme without loss of generality. The restriction is also necessary for the remaining definitions.} and can query to $\mathcal{O}_{Encrypt}$ at arbitrary times.
Here, when $\mathcal{A}$ queries to $\mathcal{O}_{Extract}$, it imposes the constraint $S\subseteq [1,n]\setminus \{i^*\}$.

\item Guess:
$\mathcal{A}$ declares a challenge keyword $w_{l^*}$ and sends it to $\mathcal{C}$.
$\mathcal{C}$ randomly chooses $\theta \in \{0,1\}$.
If $\theta=0$, then $\mathcal{C}$ sets $w_\theta=w_{l^*}$. 
Otherwise, i.e., $\theta=1$, $\mathcal{C}$ sets a random keyword as $w_\theta$, where $|w_0|=|w_1|$. 
$\mathcal{C}$ sends $c_{i^*,\theta}\gets Encrypt(params,sk,i^*,w_\theta)$ to $\mathcal{A}$.
$\mathcal{A}$ then selects $\theta'\in \{0,1\}$.
\end{itemize}

We say KASE satisfies the $(\epsilon,n)$-keyword privacy if the following relation holds for $\mathcal{A}$'s advantage with any probabilistic polynomial time algorithm and $1^\lambda$ with any sufficiently large size: 
\[
Adv:=|Pr[\theta=\theta']-1/2|<\epsilon
\]
\end{defi}

\begin{defi}[$(\epsilon,n)$-Aggregate Key Unforgeability]
In this game, an adversary $\mathcal{A}$ tries to forge a valid aggregate key where $\mathcal{A}$ can search encrypted documents with the aggregate key. 
\begin{itemize}
\item Setup:
$\mathcal{C}$ randomly chooses $i^*\in [1,n]$.
$\mathcal{C}$ generates $params\gets Setup(1^\lambda,n)$ and $sk\gets KeyGen (params)$, and then sends $params,i^*$ to $\mathcal{A}$.

\item Query:
$\mathcal{A}$ can query to $\mathcal{O}_{Extract}$ at most $n-1$ times and can query to $\mathcal{O}_{Encrypt}$ at arbitrary times. 
Here, when $\mathcal{A}$ queries to $\mathcal{O}_{Extract}$, it imposes the constraint $i^* \not\in S$.

\item Forge:
$\mathcal{A}$ outputs $S^*\subseteq [1,n]$ and $k_{agg}^*$, 
where $S^*$ includes $i^*$, i.e., $i^*\in S^*$. 
\end{itemize}

We say KASE satisfies the $(\epsilon,n)$-aggregate key unforgeability if the following relation holds for $\mathcal{A}$'s advantage with any probabilistic polynomial time algorithm, keyword $w_l$ and $1^\lambda$ with any sufficiently large size: 
\begin{equation}
\begin{array}{l}
Adv:=Pr[Test(params,Adjust(params,i^*,S^*, \nonumber \\
Trapdoor(params,k^*_{agg},S^*,w_l)))=Test(params, \nonumber \\
Adjust(params,i^*,S^*,Trapdoor(params,Extract\nonumber \\
(params, sk,S^*),S^*,w_l)))]<\epsilon \nonumber 
\end{array}
\end{equation}
\end{defi}

\begin{defi}[$(\epsilon,n)$-Trapdoor Privacy]
In this game, an adversary $\mathcal{A}$ tries to distinguish a challenge keyword or a random keyword from the given challenge trapdoor. 
\begin{itemize}
\item Init:
$\mathcal{A}$ declares a set $S^* \subseteq [1,n]$ of indexes and a challenge keyword $w_{l^*}$ used in the guess phase, and sends it to $\mathcal{C}$. 

\item Setup:
$\mathcal{C}$ generates $params\gets Setup(1^\lambda,n)$ and $sk\gets KeyGen (params)$, and then sends $params$ to $\mathcal{A}$.

\item Query:
$\mathcal{A}$ can query to $\mathcal{O}_{Trapdoor}$ at most $n-|S^*|$ times and can query to $\mathcal{O}_{Encrypt}$ at arbitrary times. 
Here, when $\mathcal{A}$ queries to $\mathcal{O}_{Encrypt}$, it imposes the constraint $w_l \not= w_{l^*} \land i \not\in S^*$. 

\item Guess:
$\mathcal{C}$ randomly chooses $\theta \in \{0,1\}$.
If $\theta=0$ then $\mathcal{C}$ sets $w_\theta=w_{l^*}$. 
Otherwise, i.e., $\theta=1$, $\mathcal{C}$ sets a random keyword as $w_\theta$, where $|w_0|=|w_1|$ holds. 
$\mathcal{C}$ sends $Tr^*\gets Trapdoor(params,k_{agg}^*,S^*,w_{\theta})$ to $\mathcal{A}$.
$\mathcal{A}$ then selects $\theta'\in \{0,1\}$.
\end{itemize}

We say KASE satisfies the $(\epsilon,n)$-trapdoor privacy if the following relation holds for $\mathcal{A}$'s advantage with any probabilistic polynomial time algorithm and $1^\lambda$ with any sufficiently large size: 
\[
Adv:=|Pr[\theta=\theta' ]-1/2|<\epsilon 
\]
\end{defi}

\subsection{Technical Difficulty}
\label{sec:difficulty}
Although one might think that the conventional searchable encryption~\cite{song2000practical,curtmola2011searchable,kamara2013parallel} can perform the same search as KASE, the number of secret keys possessed to a data user and the number of trapdoors are proportional to the number of documents stored in cloud.
Thus, the compactness cannot be satisfied.

The intuition of KASE's difficulty lies in the trade-off between security and features.
That is, there is a possibility that security cannot be satisfied if the focus is only satisfying the compactness.
In the case of KASE, the algebraic structure is limited because the sizes of aggregate key and trapdoor need to be $\mathcal{O}(1)$  regardless of the number of documents and the number of users.
Therefore, potential configurations that satisfy searchability are limited, and it has been shown that the KASE scheme of Cui et al. does not satisfy keyword privacy~\cite{kiayias2016efficient}.
Furthermore, for the trapdoor privacy, the aggregate key must already have a concrete algebraic structure, and thus a trapdoor that uses only the aggregate key necessarily has more restrictive algebraic structures. 
This makes the trapdoor privacy even more difficult to satisfy. The KASE scheme of Cui et al. also does not satisfy the trapdoor privacy (see the Appendix for details).

\section{Constructions}
\label{sec:constructions}
In this section, we propose two constructions, namely, the first construction and the main construction, following the security requirements described in Section~\ref{security_definitions}.
The first construction satisfies the keyword privacy and the aggregate key unforgeability.
The main construction satisfies the trapdoor privacy in addition to the keyword privacy and the aggregate key unforgeability.

\subsection{Idea}
Our main idea of KASE is to combine broadcast encryption (BE)~\cite{boneh2005collusion,fiat1993broadcast} and aggregate signatures (AS)~\cite{boneh2003aggregate}. BE is an encryption scheme that allows a specified set of users to decrypt ciphertext whereby a set of indexes corresponding to the user index is embedded in ciphertexts. Intuitively, the searchability can be realized by treating a decryption algorithm of BE as a test algorithm of KASE. 
The keyword privacy can then be satisfied by utilizing the ciphertext security of BE. 
Furthermore, we construct an aggregate key in the form of AS~\cite{boneh2003aggregate}. 
The signature size can be aggregated to a fixed length regardless of the number of users, and hence the compactness can be satisfied by keeping the construction of AS in aggregate keys. 
This also implies that the aggregate key unforgeability can be satisfied via the unforgeability of AS. 

When we considered the idea of combining BE and AS, we found that the BE proposed by Boneh et al.~\cite{boneh2005collusion} and the AS proposed by Boneh et al.~\cite{boneh2003aggregate} could be combined. 
Our first construction is close to a simple combination and can be constructed by a single server. However, trapdoor is out of the scope in this construction, i.e., the trapdoor privacy is unsatisfied.

For this reason, we also propose the main construction that satisfies the trapdoor privacy.
In our main construction, we embed random values in trapdoors to make the trapdoors probabilistic. In doing so, to satisfy the searchability with the trapdoors, it is necessary to embed the same random values in encrypted keywords on a cloud. However, if the random values are sent to the cloud, the cloud can also extract the original keywords from the given trapdoor. Thus, we further utilize the idea of secret sharing for these random values. 
We also prepare for two servers that do not collude with each other. 
A data user distributes the random values embedded in the trapdoors into two shares, and then sends the shares to each server individually. 
By constructing the test algorithm in a way such as $n$-$out$-$of$-$n$ threshold decryption~\cite{desmedt1994threshold,boneh2006chosen}, the random values can be embedded in encrypted keywords and public parameters without knowing the original random values themselves. The approach described above allows search in ciphertexts and satisfies the trapdoor privacy.


\subsection{First Construction}
\label{sec:first_construction}
The algorithms for the first construction are as follows:

\begin{itemize}
\item $params\gets Setup(1^\lambda,n)$:
Generate $\mathbb{B}=(p,\mathbb{G},\mathbb{G}_T,e(\cdot ,\cdot))$ as a bilinear map and bilinear groups, where $p$ is an order such that $\mathbb{G}$ and $2^\lambda < p < 2^{\lambda +1}$. 
Set $n$ as the maximum number of documents. 
For $i\in\{1,2,...,n,n+2,...,2n\}$, pick a random generator $g \in \mathbb{G}$ and a random $\alpha \in \mathbb{Z}_p$, and then compute $g_i = g^{(\alpha^i)}\in \mathbb{G}$.
Select a one-way hash function $H:\{0,1\}^* \to \mathbb{G}$.
Finally, output a public parameter $params = (\mathbb{B}, PubK, H)$, where $PubK=(g,g_1,...,g_n,g_{n+2},...,g_{2n})\in \mathbb{G}^{2n}$. 

\item $sk\gets KeyGen(params)$:
Pick a random $\beta \in \mathbb{Z}_p$ and output a secret key $sk=\beta$. 

\item $c_{i,l}\gets Encrypt(params,sk,i,w_l)$:
Pick a random $t_{i,l} \in \mathbb{Z}_p$ and output an encrypted keyword $c_{i,l}=(c_{1,i,l},c_{2,i,l},c_{3,i,l})$ by computing the following: 
\begin{eqnarray*}
c_{1,i,l}=g^{t_{i,l}} ,
c_{2,i,l}=(g^\beta \cdot g_i)^{t_{i,l}} ,
c_{3,i,l}=\frac{e(H(w_l),g)^{t_{i,l}}}{e(g_1,g_n)^{t_{i,l}}} . 
\end{eqnarray*}
\item $k_{agg}\gets Extract(params,sk,S)$:
For the given subset $S \subseteq [1,n]$ which contains the indexes of documents, 
output an aggregate key $k_{agg}$ by computing the following:
\[
k_{agg}=\Pi_{j\in S}g_{n+1-j}^{\beta}. 
\]
\item $Tr\gets Trapdoor(params,k_{agg},S,w_l)$:
For all documents relevant to the given aggregate key $k_{agg}$, 
generate a single trapdoor $Tr$ for the keyword $w_l$ by computing the following:
\[
Tr=k_{agg} \cdot H(w_l). 
\]
\item $Tr_i\gets Adjust(params,i,S,Tr)$:
For each document in the given set $S$, output trapdoor $Tr_i$ by computing the following: 
\[
Tr_{i}=Tr \cdot \Pi_{j\in S, j \not= i}g_{n+1-j+i}. 
\]
\item $b\gets Test(params,Tr_i,S,c_{i,l})$:
For the $i$th document and the keyword embedded in $Tr_i$, output $true$ or $false$ by judging whether the following equation holds or not:
\[
\frac{e(Tr_{i},c_{1,i,l})}{e(c_{2,i,l},pub)}=^? c_{3,i,l},
\]
where $pub = \Pi_{j\in S}g_{n+1-j}$. 
\end{itemize}

The first construction described above satisfies the correctness, i.e., the searchability, because equation~(\ref{first_construction_correctness}) holds.

\begin{table*}[tbp] \centering
\begin{eqnarray}
&& \frac{e(Tr_{i},c_{1,i,l})}{e(c_{2,i,l},pub)} \nonumber \\
 &=& \frac{e(\Pi_{j\in S}g_{n+1-j}^{\beta} \cdot H(w_l) \cdot \Pi_{j\in S, j \not= i}g_{n+1-j+i},g^{t_{i,l}})}{e((g^{\beta}\cdot g_i)^{t_{i,l}},\Pi_{j\in S}g_{n+1-j})} 
  = \frac{e(\Pi_{j\in S}g_{n+1-j}^{\beta} ,g^{t_{i,l}}) \cdot
e(H(w_l),g^{t_{i,l}}) \cdot
e(\Pi_{j\in S, j \not= i}g_{n+1-j+i},g^{t_{i,l}})}
{e(g^{\beta t_{i,l}},\Pi_{j\in S}g_{n+1-j}) \cdot
e(g_i^{t_{i,l}},\Pi_{j\in S}g_{n+1-j})} \nonumber \\
&=& \frac{e(\Pi_{j\in S}g_{n+1-j} ,g^{\beta t_{i,l}}) \cdot
e(H(w_l),g^{t_{i,l}}) \cdot
e(\Pi_{j\in S, j \not= i}g_{n+1-j+i},g^{t_{i,l}})}
{e(\Pi_{j\in S}g_{n+1-j},g^{\beta t_{i,l}}) \cdot
e(\Pi_{j\in S}g_{n+1-j+i}g^{t_{i,l}})} 
= \frac{e(H(w_l),g^{t_{i,l}}) \cdot
e(\Pi_{j\in S, j \not= i}g_{n+1-j+i},g^{t_{i,l}})}
{e(\Pi_{j\in S}g_{n+1-j+i},g^{t_{i,l}})} \nonumber 	\\
&=& \frac{e(H(w_l),g^{t_{i,l}}) \cdot
e(\Pi_{j\in S}g_{n+1-j+i},g^{t_{i,l}})}
{e(\Pi_{j\in S}g_{n+1-j+i},g^{t_{i,l}}) \cdot
e(g_{n+1}.g^{t_{i,l}})} 
= \frac{e(H(w_l),g^{t_{i,l}})}
{e(g_{n+1}.g^{t_{i,l}})} 
= \frac{e(H(w_l),g)^{t_{i,l}}}
{e(g_{1}.g_n)^{t_{i,l}}} .  
\label{first_construction_correctness}
\end{eqnarray}
\end{table*}
Furthermore, the sizes of an aggregate key and a trapdoor are $|\mathbb{G}|$ independent of the number of indexes in $S$ because $k_{agg}=\Pi_{j\in S}g_{n+1-j}^\beta \in \mathbb{G}$ and $Tr=k_{agg}\cdot H(w_l) \in \mathbb{G}$. 
A search over encrypted keywords can be executed without changing the encrypted keywords themselves for any keyword. Hence, the first construction satisfies the compactness. 


We will show that the first construction satisfies the keyword privacy and the aggregate key unforgeability in Section~\ref{Proof of First Construction}. 
The first construction does not satisfy the trapdoor privacy because trapdoors are deterministic with respect to keywords. In other words, an adversary can extract keywords from trapdoors when the keywords used-so-far are sent again. In the next subsection, we will introduce the main construction, which satisfies the trapdoor privacy, under a two-server setting.


\subsection{Main Construction}
\label{main_construction}

To construct a scheme that satisfies the trapdoor privacy, random values should be embedded in trapdoors. Likewise, the same random values should be embedded in the encrypted keyword and public parameters to satisfy the correctness. 


Intuitively, the approach to use random values in trapdoors seems to require a data user to send the same random values to a cloud. However, if the data user sends the same random value as those in trapdoors, the cloud can extract $k_{agg}\cdot H(w_l)$ from the given trapdoors and the random values. Consequently, the trapdoor privacy is unsatisfied. 

To overcome this limitation, instead of sending random values to a cloud, we aim to send the random values in a manner that nobody except for a data user itself can extract. 
Simultaneously, embedding the random values in encrypted keywords and a public parameter on the cloud without revealing the random values themselves. To do this, we construct trapdoors by using secret sharing. Consider a data user that distributes random values into two shares and then sends the shares to two servers. By utilizing these shares, a cloud can embed the random values in encrypted keywords and a public parameter without knowing the random values themselves. 

In the main construction that will be described below, two servers $\mathcal{C}_{main}$ and $\mathcal{C}_{aid}$ are assumed to be semi-honest and to not collude with each other without loss of generality. Both servers store the same encrypted keywords. 
When a data user generates a trapdoor, he/she also generates a random value $r$ and embeds $r$ in the resulting trapdoor. The data user then distributes $r$ into two shares and sends either of the shares to $\mathcal{C}_{main}$ and $ \mathcal{C}_{aid}$, respectively. 
After receiving the trapdoor and the share, $\mathcal{C}_{main}$ and $\mathcal{C}_{aid}$ embed the received share in the stored encrypted keyword and public parameter provisionally. Then, these values are gathered on the $\mathcal{C}_{main}$ side. 
$\mathcal{C}_{main}$ then combines the encrypted keywords and the public parameter with the shares to recover $r$. 
That is, $\mathcal{C}_{main}$ can obtain the encrypted keywords and the public parameter where the random $r$ is embedded without knowing $r$ itself. 
Consequently, $\mathcal{C}_{main}$ is able to search in encrypted files and return the search results to the data user.


The main construction is described as follows. The asymmetric bilinear map is used in the main construction to satisfy the trapdoor privacy.
Furthermore, in the Adjust and Test algorithms, we instantiate a function $f:\mathbb{Z}_p \times \mathbb{G} \to \mathbb{G}$ and $f_T:\mathbb{Z}_p \times \mathbb{G}_T \to \mathbb{G}_T$ as auxiliary functions of the input defined in Section~\ref{algorithms}. 
$f$ is a function that takes two arbitrary inputs $x\in \mathbb{Z}_p, g\in \mathbb{G}$ and outputs $g^x\in \mathbb{G}$.
$f_T$ is a function that takes two arbitrary inputs $x\in \mathbb{Z}_p, g_T\in \mathbb{G}_T$ and outputs $g_T^x\in \mathbb{G}_T$.


\begin{itemize}
\item $params\gets Setup(1^\lambda,n)$:
Generate a bilinear map group system $\mathbb{B}=(p,\mathbb{G},\mathbb{H},\mathbb{G}_T,e(\cdot ,\cdot))$, where $p$ is the order of $\mathbb{G},\mathbb{H}$ and $2^\lambda < p < 2^{\lambda +1}$.
Set $n$ as the maximum possible number of documents that belong to a data owner.
Pick a generator $g \in \mathbb{G}, h \in \mathbb{H}$ and a random $\alpha \in \mathbb{Z}_p$, and then compute $g_i = g^{(\alpha^i)}\in \mathbb{G}$ for $i\in\{1,2,...,n,n+2,...,2n\}, h_i = h^{(\alpha^i)}\in \mathbb{H}$ for $i\in[1,n]$.
Select a one-way hash function $H:\{0,1\}^* \to \mathbb{G}$.
Finally, output a public parameter $params = (\mathbb{B}, PubK, H)$, where $PubK=(g,g_1,...,g_n,g_{n+2},...,g_{2n},h,h_1,...,h_n)\in (\mathbb{G}^{2n}\times \mathbb{H}^{n+1})$.

\item $sk\gets KeyGen(params)$:
Pick a random $\beta \in \mathbb{Z}_p$ and output a secret key $sk=\beta$. 

\item $c_{i,l}\gets Encrypt(params,sk,i,w_l)$:
Pick a random $t_{i,l} \in \mathbb{Z}_p$ and output an encrypted keyword $c_{i,l}=(c_{1,i,l},c_{2,i,l},c_{3,i,l})$ by computing the following: 
\begin{equation}
\begin{array}{l}
c_{1,i,l}=h^{t_{i,l}} \in \mathbb{H}, \\
c_{2,i,l}=(g^\beta \cdot g_i)^{t_{i,l}} \in \mathbb{G}, \\
c_{3,i,l}= \frac{e(H(w_l),h)^{t_{i,l}}}{e(g_1,h_n)^{t_{i,l}}} \in \mathbb{G}_T.  \nonumber
\end{array}
\end{equation}

\item $k_{agg}\gets Extract(params,sk,S)$:
For the given subset $S \subseteq [1,n]$ which contains the indexes of documents, 
output an aggregate key $k_{agg}$ by computing the following:
\[
k_{agg}=\Pi_{j\in S}g_{n+1-j}^{\beta}\in \mathbb{G}. 
\]

\item $Tr\gets Trapdoor(params,k_{agg},S,w_l)$:
Randomly generate $r \in \mathbb{Z}_p$ and calculate $Tr=(k_{agg} \cdot H(w_l))^r\in \mathbb{G}$.
Then, $r$ is broken into $r=r_{main} + r_{aid}$, and $Tr_{main}=(Tr,r_{main}),Tr_{aid}=r_{aid}$.

\item $Tr_{i}\gets Adjust(params,i,S,Tr,f(r_{main},pub_i),f(r_{aid},\\pub_i))$:
Calculate $pub_i=\Pi_{j\in S, j \not= i}g_{n+1-j+i}\in \mathbb{G}$ on the two servers.
$\mathcal{C}_{aid}$ sends $f(r_{aid},pub_i)=pub_i^{r_{aid}}$ to $\mathcal{C}_{main}$.
Next, $\mathcal{C}_{main}$ calculates $(f(r_{main},pub_i))\cdot (f(r_{aid},pub_i))=pub_i^{r_{main}}\cdot pub_i^{r_{aid}}=pub_i^r$ and calculates $Tr_i=Tr\cdot pub_i^r\in \mathbb{G}$.

\item $b\gets Test(params,Tr_i,S,c_{i,l},f_T(r_{main},c^\#_{2,i,l}),f_T(r_{aid},\\c^\#_{2,i,l}),f_T(r_{main},c_{3,i,l}),f_T(r_{aid},c_{3,i,l}))$:
Calculate $pub=\Pi_{j\in S}h_{n+1-j}\in \mathbb{H}$ and $c^\#_{2,i,l}=e(c_{2,i,l},pub)$ on the two servers.
$\mathcal{C}_{aid}$ sends $f_T(r_{aid},c^\#_{2,i,l})=e(c_{2,i,l},pub)^{r_{aid}}$ to $\mathcal{C}_{main}$.
Next, $\mathcal{C}_{main}$ calculates $(f_T(r_{main},c^\#_{2,i,l}))\cdot (f_T(r_{aid},c^\#_{2,i,l}))=e(c_{2,i,l},pub)^{r_{main}}\cdot e(c_{2,i,l},pub)^{r_{aid}}=e(c_{2,i,l},pub)^r$.
In addition, $\mathcal{C}_{aid}$ sends $f_T(r_{aid},c_{3,i,l})=c_{3,i,l}^{r_{aid}}$ to $\mathcal{C}_{main}$.
Then, $\mathcal{C}_{main}$ calculates $(f_T(r_{main},c_{3,i,l}))\cdot (f_T(r_{aid},c_{3,i,l}))=c_{3,i,l}^{r_{main}}\cdot c_{3,i,l}^{r_{aid}}=c_{3,i,l}^r$ and outputs $true$ or $false$ by judging whether the following equation holds or not:
\[
\frac{e(Tr_{i},c_{1,i,l})}{e(c_{2,i,l},pub)^r}=^? c_{3,i,l}^r .
\]
\end{itemize}

The main construction satisfies the correctness, i.e., the searchability, as shown in equation~(\ref{main_construction_correctness}). 

\begin{table*}[tbp] \centering
\begin{eqnarray}
&& \frac{e(Tr_{i},c_{1,i,l})}{e(c_{2,i,l},pub)^r} \nonumber \\
&=&\frac{e(\Pi_{j\in S}g_{n+1-j}^{\beta r} \cdot H(w_l)^r \cdot \Pi_{j\in S, j \not= i}g_{n+1-j+i}^r,h^{t_{i,l}})}{e((g^{\beta}\cdot g_i)^{t_{i,l}},\Pi_{j\in S}h_{n+1-j}^{r})} 
=\frac{e(\Pi_{j\in S}g_{n+1-j}^{\beta r} ,h^{t_{i,l}}) \cdot
e(H(w_l)^r,h^{t_{i,l}}) \cdot
e(\Pi_{j\in S, j \not= i}g_{n+1-j+i}^r,h^{t_{i,l}})}
{e(g^{\beta t_{i,l}},\Pi_{j\in S}h_{n+1-j}^{r}) \cdot
e(g_i^{t_{i,l}},\Pi_{j\in S}h_{n+1-j}^{r})} \nonumber \\
&=&\frac{e(\Pi_{j\in S}g_{n+1-j}^{r} ,h^{\beta t_{i,l}}) \cdot
e(H(w_l)^r,h^{t_{i,l}}) \cdot
e(\Pi_{j\in S, j \not= i}g_{n+1-j+i}^r,h^{t_{i,l}})}
{e(\Pi_{j\in S}g_{n+1-j}^{r},h^{\beta t_{i,l}}) \cdot
e(\Pi_{j\in S}g_{n+1-j+i}^{r},h^{t_{i,l}})}
=\frac{e(H(w_l)^r,h^{t_{i,l}}) \cdot
e(\Pi_{j\in S, j \not= i}g_{n+1-j+i}^r,h^{t_{i,l}})}
{e(\Pi_{j\in S}g_{n+1-j+i}^{r},h^{t_{i,l}})} \nonumber \\
&=&\frac{e(H(w_l)^r,h^{t_{i,l}}) \cdot
e(\Pi_{j\in S}g_{n+1-j+i}^r,h^{t_{i,l}})}
{e(\Pi_{j\in S}g_{n+1-j+i}^{r},h^{t_{i,l}}) \cdot
e(g^r_{n+1}.h^{t_{i,l}})}
=\frac{e(H(w_l)^r,h^{t_{i,l}})}
{e(g^r_{n+1},h^{t_{i,l}})}
=\frac{e(H(w_l),h)^{r t_{i,l}}}
{e(g_{1}.h_n)^{r t_{i,l}}}
=c^r_{3,i,l}
\label{main_construction_correctness}
\end{eqnarray}
\end{table*}

Furthermore, similar to the first construction, the sizes of an aggregate key and a trapdoor are $|\mathbb{G}|$ independent of the number $n$ of indexes in $S$. In addition, the data size of encrypted keywords is independent of the number of users because the algorithms do not change the encrypted keywords themselves. 
Thus, the main construction also satisfies the compactness.

\section{Security Proofs} \label{Proofs}
In this section, we will show the security proofs of the first construction and main construction. The security of the main construction is proved in the two-server setting because it uses to servers. 
The proof statement is consistent with the security definitions because our definitions have captured the multi-server setting by applying an auxiliary function individually for each server.

\subsection{Proofs of the First Construction} \label{Proof of First Construction}
The first construction satisfies the $(\epsilon',n)$-keyword privacy and the $(\epsilon',n)$-aggregate key unforgeability. 
In this section, we prove these two securities.

\begin{theo}[$(\epsilon',n)$-Keyword Privacy]
The first construction satisfies the $(\epsilon',n)$-keyword privacy under the $(\epsilon,n)$-BDHE Assumption, where $\epsilon\geq\epsilon'$.
\end{theo}

$Proof.$
Suppose there exists an adversary $\mathcal{A}$, whose advantage is $\epsilon'$, against the first construction.
We then build an algorithm $\mathcal{B}$ that solves the BDHE problem. 
Let $\mathcal{C}$ be a challenger for the BDHE problem. 
Algorithm $\mathcal{B}$ proceeds as follows.

\begin{itemize}
\item Init:
$\mathcal{A}$ declares challenge file index $i^* \in [1,n]$ and sends it to $\mathcal{C}$.
\item Setup:
$\mathcal{C}$ sends $(g,h,g_1,g_2,...,g_n,g_{n+2},...,g_{2n},Z)$ to $\mathcal{B}$.
$\mathcal{B}$ randomly generates $sk=\beta$ and calculates $v'=g^\beta g_{i^*}^{-1}$.
$\mathcal{B}$ sends $params=(g,g_1,g_2,...,g_n,g_{n+2},...,g_{2n})$ to $\mathcal{A}$.
\item Query:
When $\mathcal{A}$ queries for $\mathcal{O}_{Extract}$, $\mathcal{B}$ responds as follows:
\begin{itemize}
\item
If an aggregate key for $i^* \in S$ is queried, return $\bot$.
\item
If an aggregate key for $i^* \not\in S$ is queried, return $k_{agg}=(\Pi_{j\in S}g_{n+1-j}^{\beta}) \cdot (\Pi_{j\in S}g_{n+1-j+i^*})^{-1}=\Pi_{j\in S}g_{n+1-j}^{\beta - \alpha^{i^*}}$.
If $j=i^*$, $(\Pi_{j\in S}g_{n+1-j+i^*})^{-1}$ cannot be calculated, but it can be calculated because of $i^* \not\in S$.
\end{itemize}

When $\mathcal{A}$ queries for $\mathcal{O}_{Encrypt}$, $\mathcal{B}$ randomly generates $t_{i,l}\in \mathbb{Z}_p^*$, calculates the following $c_{i,l}=(c_{1,i,l},c_{2,i,l},c_{3,i,l})$ and responds to $\mathcal{A}$ ($c_{1,i,l}=g^{t_{i,l}},c_{2,i,l}=(v' \cdot g_i)^{t_{i,l}} ,c_{3,i,l}=\frac{e(H(w_{l}),g)^{t_{i,l}}}{e(g_1,g_n)^{t_{i,l}}}$).

\item Guess:
$\mathcal{A}$ declares the challenge keyword $w_{l^*}$ and sends it to $\mathcal{B}$.
$\mathcal{B}$ calculates the challenge encrypted keyword $c_{1,i^*,\theta}=h,c_{2,i^*,\theta}=h^\beta ,c_{3,i^*,\theta}=\frac{e(H(w_{l^*}),h)}{Z}$.

Here, we define $h=g^t$ ($t$ is a random value).
Then, when $Z=e(g_{n+1},h)$, $c_{1,i^*,\theta}=g^t=h ,c_{2,i^*,\theta}=((g^\beta g_{i^*}^{-1})\cdot g_{i^*})^t=g^{\beta t}=h^\beta, c_{3,i^*,\theta}=\frac{e(H(w_{l^*}),g)^t}{e(g_1,g_n)^t}=\frac{e(H(w_{l^*}),h)}{e(g_{n+1},h)}=\frac{e(H(w_{l^*}),h)}{Z}$.
Therefore, the calculation results are identical to the results of the Encrypt algorithm of the first construction. 
$\mathcal{B}$ sends $c_{i^*,\theta}=(c_{1,i^*,\theta},c_{2,i^*,\theta},c_{3,i^*,\theta})$ to $\mathcal{A}$.
$\mathcal{A}$ chooses $\theta'\in \{0,1\}$ and sends it to $\mathcal{B}$.
Then, $\mathcal{B}$ sends $\theta'$ to $\mathcal{C}$ as a guess of $\theta$.
\end{itemize}
In the guess phase, if $Z$ is a random value, then $\Pr[\theta=\theta']=1/2$. 
On the other hand, if $Z=e(g_{n+1},h)$, then $|\Pr[\theta=\theta']-1/2|>\epsilon'$. 
This indicates that $\mathcal{B}$ has an advantage over $\epsilon'$ for solving the $(\epsilon,n)$-BDHE problem. 
Thus, if the $(\epsilon,n)$-BDHE assumption holds, the first construction satisfies the $(\epsilon',n)$-keyword privacy.


\begin{theo}[$(\epsilon',n)$-Aggregate key Unforgeability]
The first construction satisfies the $(\epsilon',n)$-aggregate key unforgeability under the $(\epsilon,n)$-DHE Assumption, where $\epsilon=\epsilon'$.
\end{theo}

$Proof.$
Suppose there exists as adversary $\mathcal{A}$, whose advantage is $\epsilon'$, against the first construction.
We then build an algorithm $\mathcal{B}$ that solves the DHE problem. 
Let $\mathcal{C}$ be a challenger for the DHE problem. 
Algorithm $\mathcal{B}$ proceeds as follows.

\begin{itemize}
\item Setup:
$\mathcal{C}$ sends $(g,g^{\alpha},g^{\alpha^2},...,g^{\alpha^n},g^{\alpha^{n+2}},...,g^{\alpha^{2n}})$ to $\mathcal{B}$.
$\mathcal{B}$ randomly generates $sk=\beta$ and calculates $v'=g^\beta g_{i^*}^{-1}$.
$\mathcal{B}$ sends $params=(g,g_1,g_2,...,g_n,g_{n+2},...,g_{2n})$ to $\mathcal{A}$.
\item Query:
When $\mathcal{A}$ queries for $\mathcal{O}_{Extract}$, $\mathcal{B}$ responds as follows:
\begin{itemize}
\item
If an aggregate key for $i^* \in S$ is queried, return $\bot$.
\item
If an aggregate key for $i^* \not\in S$ is queried, return $k_{agg}=(\Pi_{j\in S}g_{n+1-j}^{\beta}) \cdot (\Pi_{j\in S}g_{n+1-j+i^*})^{-1}=\Pi_{j\in S}g_{n+1-j}^{\beta - \alpha^{i^*}}$.
Here, if $j=i^*$, then $(\Pi_{j\in S}g_{n+1-j+i^*})^{-1}$ cannot be calculated, but it can be calculated because of $i^* \not\in S$.
\end{itemize}

When $\mathcal{A}$ queries for $\mathcal{O}_{Encrypt}$, $\mathcal{B}$ randomly generates $t_{i,l}\in \mathbb{Z}_p^*$, calculates $c_{i,l}=(c_{1,i,l},c_{2,i,l},c_{3,i,l})$ and responds to $\mathcal{A}$ ($c_{1,i,l}=g^{t_{i,l}},c_{2,i,l}=(v' \cdot g_i)^{t_{i,l}} ,c_{3,i,l}=\frac{e(H(w_{l}),g)^{t_{i,l}}}{e(g_1,g_n)^{t_{i,l}}}$).

\item Forge:
$\mathcal{A}$ outputs $S^*,k_{agg}^*$ and sends them to $\mathcal{B}$.
\begin{itemize}
\item If $i^*\not\in S^*$, abort
\item If $i^*\in S^*$, $k_{agg}^*=(\Pi_{j\in S,j\not= i^*}g_{n+1-j})^{\beta -\alpha^{i^*}}\cdot (g_{n+1-i^*})^{\beta- \alpha^{i^*}}$.
By using this $k_{agg}^*$, $\mathcal{B}$ calculates (\ref{aggregate_key_forge}) and outputs results.
\begin{table*}[tbp] \centering
\begin{eqnarray}
&&\frac{(\Pi_{j\in S^*,j\not= i^*}g_{n+1-j})^\beta \cdot (\Pi_{j\in S^* ,j\not= i^*}g_{n+1-j+i^*})^{-1} \cdot (g_{n+1-i^*})^\beta}{k_{agg}^*} \nonumber \\
&=&\frac{(\Pi_{j\in S^*,j\not= i^*}g_{n+1-j})^\beta \cdot (\Pi_{j\in S^* ,j\not= i^*}g_{n+1-j+i^*})^{-1} \cdot (g_{n+1-i^*})^\beta}{(\Pi_{j\in S^*,j\not= i^*}g_{n+1-j})^\beta \cdot (\Pi_{j\in S^*,j\not= i^*}g_{n+1-j})^{-\alpha^{i^*}}\cdot (g_{n+1-i^*})^\beta \cdot (g_{n+1-i^*})^{-\alpha^{i^*}}} 
=(g_{n+1-i^*})^{\alpha^{i^*}} 
=g^{\alpha^{n+1-i^*+i^*}} 
=g^{\alpha^{n+1}}
\label{aggregate_key_forge}
\end{eqnarray}
\end{table*}
\end{itemize}
\end{itemize}
The result in the above $(\epsilon',n)$-aggregate key unforgeability game is identical to the answer of the $(\epsilon,n)$-DHE problem. 
That is, the advantage of the $(\epsilon',n)$-aggregate key unforgeability game is equal to the advantage of the $(\epsilon,n)$-DHE problem. 
Thus, if the $(\epsilon,n)$-DHE assumption holds, the first construction satisfies the $(\epsilon',n)$-aggregate key unforgeability.

\subsection{Proofs of the Main Construction}
In this section, we prove that the main construction satisfies the $(\epsilon',n)$-trapdoor privacy. 
Note that the $(\epsilon',n)$-keyword privacy and the $(\epsilon',n)$-aggregate key unforgeability can be proved similarly to the proofs for the first construction except for the use security assumptions in asymmetric bilinear groups (see the  Appendix for details). 
As described in the previous section, the main construction is based on two servers. 
In our security proof, a challenge ciphertext and a challenge trapdoor for both $\mathcal{C}_{main}$ and $\mathcal{C}_{aid}$ are simulated by a reduction algorithm. 

\begin{theo}[$(\epsilon',n)$-Trapdoor Privacy]
Let a hash function $H$ be modeled as a random oracle.
The main construction satisfies the $(\epsilon',n)$-trapdoor privacy under the $\epsilon$-XDH assumption, where $\epsilon\geq\epsilon'$.
\end{theo}

$Proof.$
Suppose there exists as adversary $\mathcal{A}$, whose advantage is $\epsilon'$, against the main construction.
We then build an algorithm $\mathcal{B}$ that solves the XDH problem. 
Let $\mathcal{C}$ be a challenger for the XDH problem. 
Algorithm $\mathcal{B}$ proceeds as follows.

\begin{itemize}
\item Init:
$\mathcal{A}$ declares challenge file index set $S^* \subseteq U$ and challenge keyword $w_{l^*}$ and sends them to $\mathcal{C}$.

\item Setup:
$\mathcal{C}$ sends $(g,h,g^a,g^b,Z)$ to $\mathcal{B}$.
$\mathcal{B}$ randomly generates $\alpha, \omega, r'_{main}\in \mathbb{Z}_p^*$ and calculates $g_i=g^{\alpha^i}(i\in \{1,2,...,n,n+2,...,2n\}), h_i=h^{\alpha^i}(i\in [1,n+1])$.
$\omega$ is used to generate both a response from the random oracle $H$ and the challenge trapdoor. Simultaneously, $r'_{main}$ corresponds to $r_{main}$ of the challenge trapdoor.
In this proof, the random value $r$ of challenge trapdoor is mapped to the challenge $a$ of $g^a$. In doing so, calculating $r'_{aid}=a-r'_{main}$ is necessary in accordance with the Trapdoor algorithm. However, since $\mathcal{B}$ does not know $a$, $\mathcal{B}$ cannot calculate $r'_{aid}$ itself. 
Then, instead of calculating $r'_{aid}$ as behavior for $\mathcal{C}_{aid}$ in the challenge phase, $\mathcal{B}$ calculates the value including $r'_{aid}$ as $\mathcal{C}_{aid}$. Specifically, $\mathcal{B}$ calculates $g^a \cdot g^{-r'_{main}}=g^{a- r'_{main}}=g^{r'_{aid}}$ and generates the value including $g^{r'_{aid}}$ as the behavior of $\mathcal{C}_{aid}$. 
This implicitly means that, in the main construction, $\mathcal{C}_{aid}$, who receives $r_{aid}$, embeds $r_{aid}$ in $e(c_{2,i,l},pub),pub_i$ and $c_{3,i,l}$ in the Adjust and Test algorithms.
To execute the Adjust and Test algorithms correctly for the challenge trapdoor, $\mathcal{C}_{aid}$ is also considered to use the following values:  
\begin{equation}
\begin{array}{l}
pub^*=\Pi_{j\in S^*}h_{n+1-j}, \\
pub^*_i=\Pi_{j\in S^*,j\not= i}(g^{r'_{aid}})^{\alpha^{n+1-j+i}}(i\in S^*). \nonumber
\end{array}
\end{equation}
Next, $\mathcal{B}$ calculates the following values:
\begin{equation}
\begin{array}{l}
c'^\#_{2,i,l}=(Z\cdot g^{b r'_{main}})\cdot (g^{r'_{aid}})^{\alpha^i}(i\in S^*), \\
A_i=e(c'^\#_{2,i,l},pub^*), \\
B=\frac{e((g^{r'_{aid}})^\omega, h)}{e(g^{r'_{aid}},h_{n+1})}=(\frac{e(g^\omega, h)}{e(g_1,h_n)})^{r'_{aid}}. \nonumber
\end{array}
\end{equation}
These values are used to calculate encrypted keywords, which satisfy the correctness for the challenge trapdoor.
$\mathcal{B}$ sends $params=(g,g_1,g_2,...,g_n,g_{n+2},...,g_{2n},h,h_1,h_2,...,h_n)$ to $\mathcal{A}$.
Here, $\mathcal{B}$ does not know $sk=b$ because it is a part of the secret of the challenge.
\item Hash:
When $\mathcal{A}$ queries for random oracle model, $\mathcal{B}$ responds as follows.
Note that $\mathcal{B}$ has a hash list $L(w_l,y_l)$.
In the initial state, $L$ is an empty set.
\begin{itemize}
\item
If $w_l$ is in $L$, return the corresponding $y_l$ stored in $L$.
\item
If $w_l$ is not in $L$, setting $y_l$ as follows, add $(w_l,y_l)$ to $L$ and return $y_l$.
\begin{itemize}
\item
If $w_l=w_{l^*}$, let $y_l=g^{\omega}$.
\item
If $w_l\not=w_{l^*}$, choose $x \in \mathbb{Z}_p^*$ uniformly randomly and let $ y_l = g ^ {x} $.
\end{itemize}
\end{itemize}

\item Query:
If $\mathcal{A}$ queries for $\mathcal{O}_{Encrypt}$, $\mathcal{B}$ returns as follows:
\begin{itemize}
\item
If $\mathcal{A}$ queries for a keyword which satisfies $w_l = w_{l^*} \lor i \in S^*$, return $\bot$.
\item
If $\mathcal{A}$ queries for a keyword which satisfies $w_l \not= w_{l^*} \land i \not\in S^*$, randomly generate $t_{i,l} \in \mathbb{Z}_p^*$ and return $c_{i,l}=(c_{1,i,l},c_{2,i,l},c_{3,i,l})$
($c_{1,i,l}=h^{t_{i,l}} ,c_{2,i,l}=(g^b \cdot g_i)^{t_{i,l}} ,c_{3,i,l}=\frac{e(H(w_l),h)^{t_{i,l}}}{e(g_1,h_n)^{t_{i,l}}}$)
\end{itemize}

If $\mathcal{A}$ queries for $\mathcal{O}_{Trapdoor}$, $\mathcal{B}$ randomly generate $r\in \mathbb{Z}_p^*$ and calculate $Tr=(\Pi_{j\in S}(g^b)^{\alpha^{n+1-j}} \cdot H(w_l))^r=(\Pi_{j\in S}g_{n+1-j}^b \cdot H(w_l))^r$.
Then, $\mathcal{B}$ randomly generates $r_{main}\in \mathbb{Z}_p^*$ and calculates $r_{aid}=r-r_{main}$.
$\mathcal{B}$ returns $(Tr,r_{main})$ to $\mathcal{A}$.

\item Guess:
$\mathcal{B}$ calculates the challenge trapdoor as $Tr^*=\Pi_{j\in S^*}Z^{\alpha^{n+1-j}} \cdot (g^a)^\omega$. 
At this time, if $Z=g^{ab}$, $Tr=\Pi_{j\in S^*}(g^{ab})^{\alpha^{n+1-j}} \cdot (g^a)^\omega=(\Pi_{j\in S^*}g^b_{n+1-j} \cdot g^\omega)^a$ holds. 
From the simulation of the hash phase, $H(w_{l^*})=g^{\omega}$ holds, and therefore $Tr=(\Pi_{j\in S^*}g^b_{n+1-j} \cdot H(w_{l^*}))^a$ holds. 
Therefore, $Tr$ has the same distribution as that calculated by the Trapdoor algorithm in the main construction. 
$\mathcal{B}$ sends $Tr^*$ and $r'_{main}$ to $\mathcal{A}$. 
$\mathcal{A}$ then chooses $\theta'\in\{0,1\}$ and returns it to $\mathcal{B}$. 
Finally, $\mathcal{B}$ returns the received $\theta'$ to $\mathcal{C}$ as a guess of $\theta$.
\end{itemize}

Note that if $Z=g^{ab}$, not only the trapdoor generated in the query phase but also the challenge trapdoor satisfies the correctness. 

In the guess phase, if $Z$ is a random value, then $Pr[\theta=\theta']=1/2$.
On the other hand, if $Z=g^{ab}$, then $|Pr[\theta=\theta']-1/2|=Adv>\epsilon'$.
This indicates that $\mathcal{B}$ has an advantage over $\epsilon'$ for solving the $\epsilon$-XDH problem.
Thus, if the $\epsilon$-XDH assumption holds, the main construction satisfies the $(\epsilon',n)$-trapdoor privacy.

\section{Discussion}
\label{Discussion}
In this section, we discuss performance of our proposed schemes. We first implement the proposed schemes to measure their actual performance. Next, we theoretically compare the computational cost and the storage cost of the proposed schemes with those of related works. We also compare the security features with related works.

\subsection{Implementation and Performance}
We implement the first construction and the main construction to evaluate the performance of each algorithm. The implementation environment and performance evaluation are as follows:

\subsubsection{Implementation Environment}
In our implementation, we use the mcl library version 0.94\footnote{mcl library: \url{ https://github.com/herumi/mcl}}, which is a C++  library for pairing computation. 
We also use the BLS12-381 curves. The curves are asymmetric bilinear maps, and hence the parameters in the first construction are dually generated for each input group of the bilinear maps. 
We also evaluate each cryptographic operation in the C++ platform with Mac OS named Mojave. In our environment, CPU is 1.4 GHz Intel Core i5-4260U and memory is 4 GB 1600 MHz DDR3. 
We note that the communication latency to interact between $\mathcal{C}_{main}$ and $\mathcal{C}_{aid}$ in the main construction is not measured because the communication process between $\mathcal{C}_{main}$ and $\mathcal{C}_{aid}$ is not implemented due to the lack of communication function in the mcl library.

\subsubsection{Performance Evaluation}
\begin{figure*}[htpb]
  \centering
    \begin{tabular}{c}
      \begin{minipage}{0.33\hsize}
        \centering
          \includegraphics[keepaspectratio, scale=0.55, angle=0]{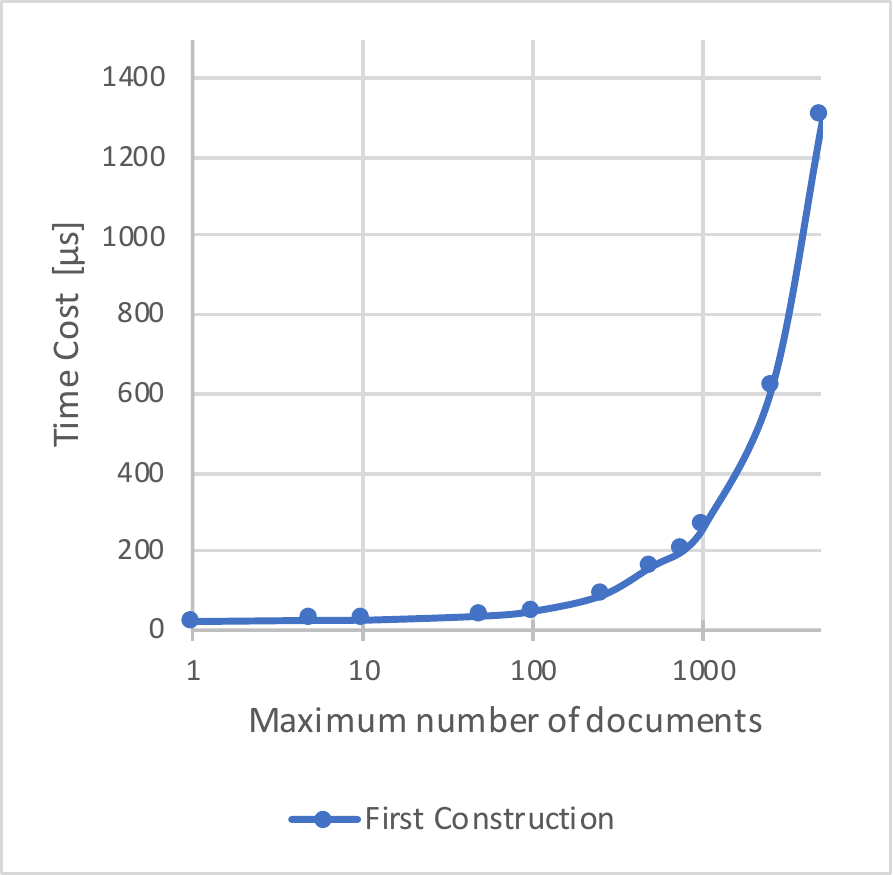}
                          \caption{Time cost of Setup}
                          \label{fig:setup}
      \end{minipage}
      \begin{minipage}{0.33\hsize}
        \centering
          \includegraphics[keepaspectratio, scale=0.55, angle=0]{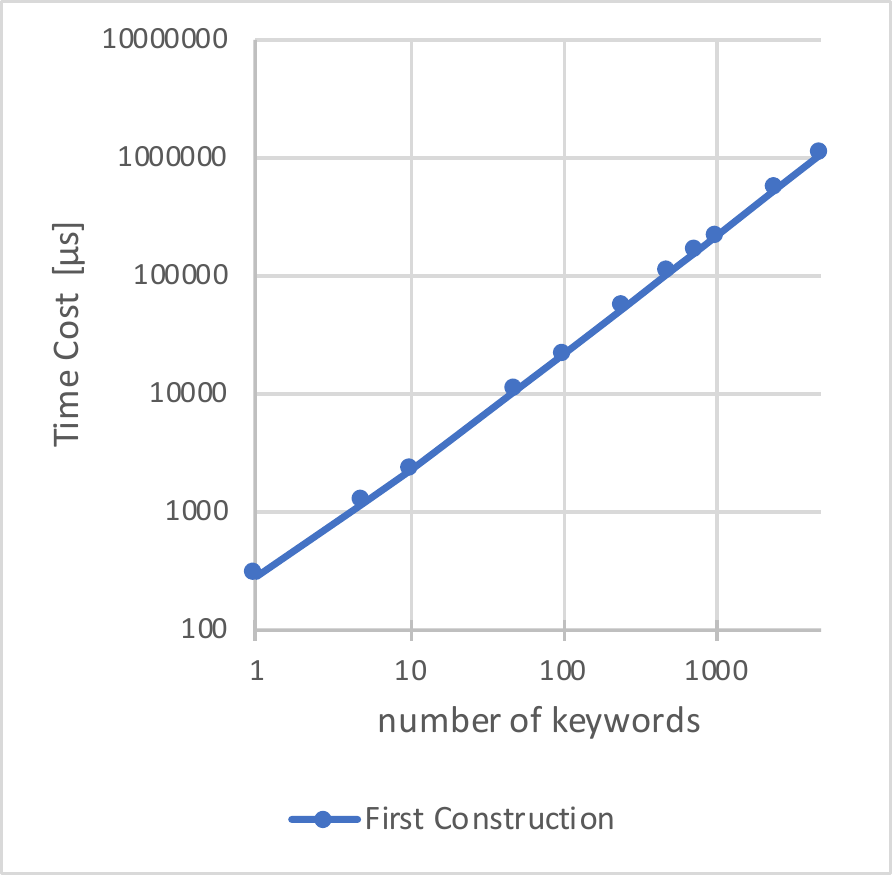}
                          \caption{Time cost of Encrypt}
                          \label{fig:encrypt}
      \end{minipage} 
      \begin{minipage}{0.33\hsize}
        \centering
          \includegraphics[keepaspectratio, scale=0.55, angle=0]
                          {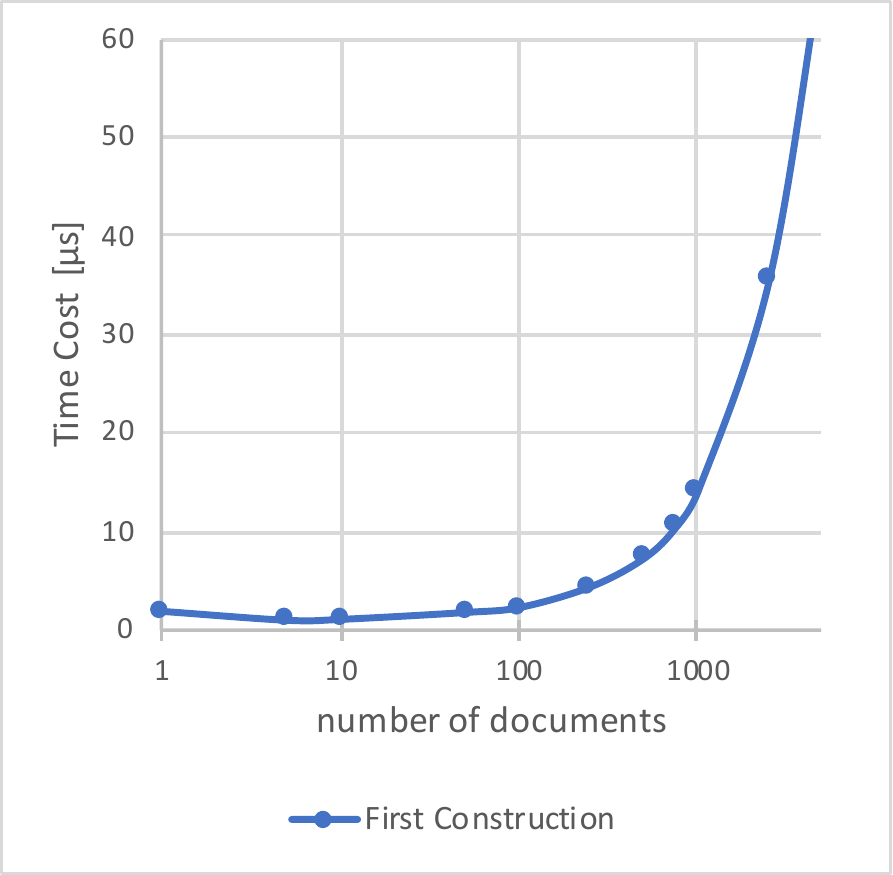}
                          \caption{Time cost of Extract}
                          \label{fig:extract}
      \end{minipage} \\
      
      \begin{minipage}{0.06\hsize}
      \vspace{7mm}
      \end{minipage} \\
      
      \begin{minipage}{0.33\hsize}
        \centering
          \includegraphics[keepaspectratio, scale=0.55, angle=0]
                          {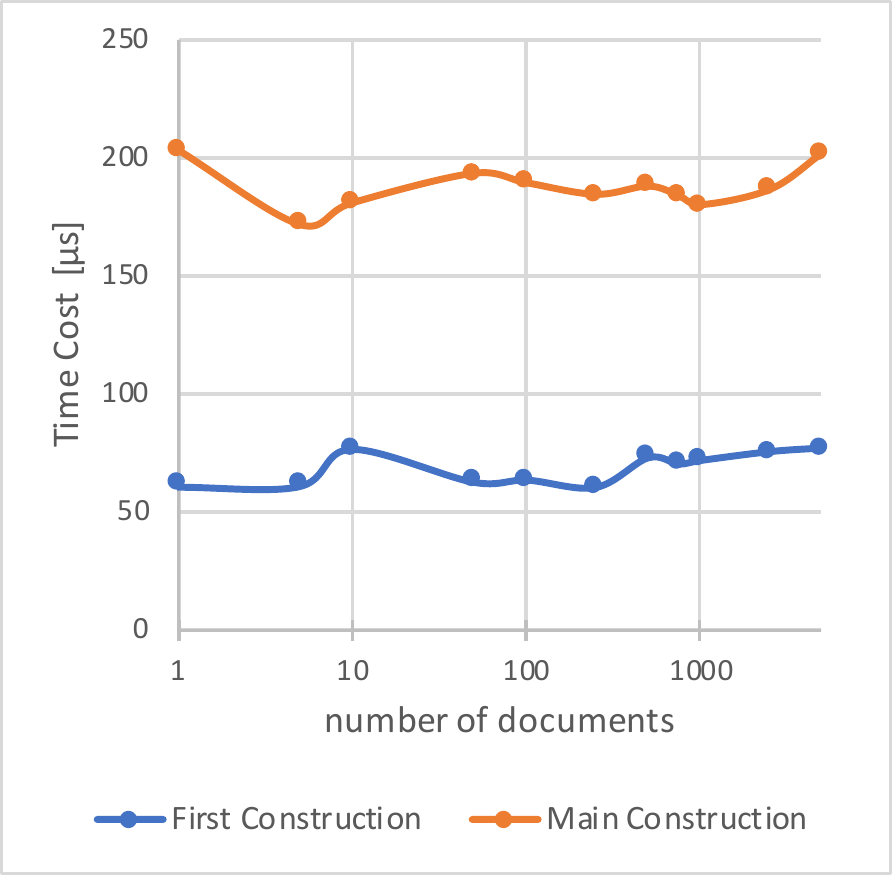}
                          \caption{Time cost of Trapdoor}
                          \label{fig:trapdoor}
      \end{minipage}
      \begin{minipage}{0.33\hsize}
        \centering
          \includegraphics[keepaspectratio, scale=0.55, angle=0]
                          {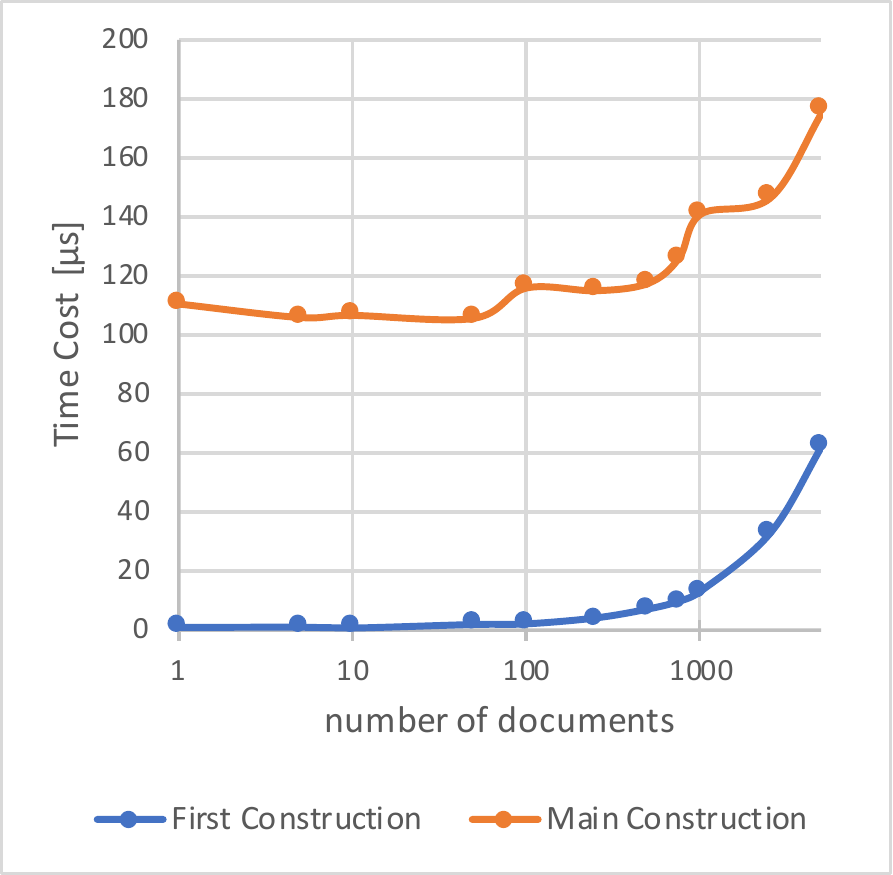}
                          \caption{Time cost of Adjust}
                          \label{fig:adjust}
      \end{minipage}
      \begin{minipage}{0.33\hsize}
        \centering
          \includegraphics[keepaspectratio, scale=0.55, angle=0]
                          {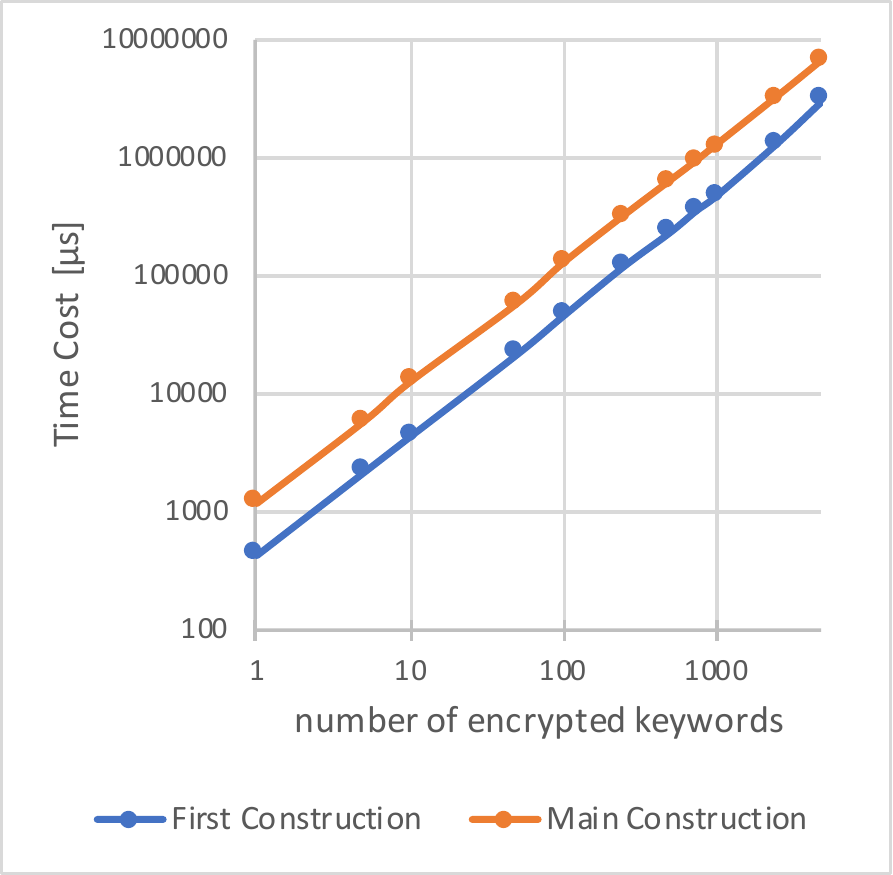}
                          \caption{Time cost of Test}
                          \label{fig:test}
      \end{minipage}
    \end{tabular}
\end{figure*}

The results of the performance evaluation of our schemes are shown in Figures~~\ref {fig:setup}-\ref{fig:test}. 
Among the KASE algorithms, the Setup, Encrypt, and Extract algorithms are common to the first construction and the main construction because the mcl library supports asymmetric bilinear maps. Thus, the evaluation of these algorithms, i.e., Figures~\ref {fig:setup}-\ref{fig:extract}, contains only the first construction. The results shown in Figures~\ref {fig:setup}-\ref{fig:extract} are common also for those in the main construction. 
On the other hand, the results of the Adjust and Test algorithms listed in Figures~\ref{fig:adjust}-\ref{fig:test} do not include the communication latency between the two servers in the main construction.

The execution time of the Setup algorithm is linear with respect to the maximum number of documents belonging to a single owner. When the number of documents increases to 5,000, the Setup algorithm only needs one millisecond of execution time and therefore remains reasonable. 

The execution time of the Encrypt algorithm is linear with respect to the number of keywords. 
The execution time of the Encrypt algorithm is larger than those of other algorithms because of the use of bilinear maps with heavy calculations. Nevertheless, when the number of documents increases to 5,000, the execution time of the Encrypt algorithm finishes within one second and is therefore still practical .

The execution time of the Extract algorithm is linear with respect to the number of shared documents. When the number of documents increases to 5,000, the Extract algorithm needs only 0.07 milliseconds of execution time. The Extract algorithm can be performed faster than the other algorithms.

The execution time of the Trapdoor algorithm is constant with respect to the number of documents, i.e., 0.07 milliseconds in the first construction and 0.2 milliseconds in the main construction. The difference in the execution time depends on the number of scalar multiplications. In particular, the first construction does not utilize scalar multiplication, whereas the main construction utilizes a single scalar multiplication. Since the data sizes of $r_{main}$ and $r_{aid}$ are small because of integers, the entire execution time can be minimized even when the communication latency is included. 

The execution time of the Adjust algorithm is linear with respect to the number of documents. 
When the number of documents increases to 5,000, the Adjust algorithm takes only 0.06 milliseconds in the first construction and only 0.18 milliseconds in the main construction. Similar to the Extract algorithm, the Adjust algorithm in the first construction can be performed faster than the other algorithms. In the main construction, the Adjust algorithm includes scalar multiplications and hence the computation is slightly heavy. However, we note that the computations for $\mathcal{C}_{main}$ and $\mathcal{C}_{aid}$ can be done in parallel to improve the performance. Although we did not implement the parallelization, this could improve the performance twice faster.


The execution time of the Test algorithm is linear with respect to the number of encrypted keywords. When the number of documents increases to 5,000, the algorithm takes three seconds in the first construction and six seconds in the main construction. Similar to the Encrypt algorithm, the high cost is caused by the use of bilinear maps, which is a bottleneck in all algorithms that use it. However, the search process can be fully parallelized because it is individual for each encrypted keyword. Thus, the performance can be improved by parallelization, e.g., by the use of the OpenMP library\footnote{OpenMP library: \url{ https://www.openmp.org/wp-content/uploads/cspec20.pdf}}. 

Finally, search for any keyword is a summation of the execution times of the Trapdoor, Adjust, and Test algorithms. For instance, a search in 5,000 encrypted keywords is executed within about three seconds in the first construction and about six seconds in the main construction. 

\subsection{Computational and Storage Cost Analysis}
In this section, we compare the computational cost and the storage cost of our schemes with other schemes~\cite{cui2016key,li2016verifiable,zhou2018file} of KASE. The results are shown in Tables \ref{tab:computation_KASE} and \ref{tab:cost_KASE}. 
Li et al.~\cite{li2016verifiable} proposed two constructions, i.e., the single-owner setting and the multi-owner setting. Therefore, we only compare our schemes in the single-owner setting because our schemes have a single-owner setting. 

\begin{table*}[htb]
\begin{center}
\begin{tabular}{c|c|c|c|c|c} \hline
& Cui et al. \cite{cui2016key} & Li et al. \cite{li2016verifiable} & Zhou et al.\cite{zhou2018file} & First Construction & Main Construction \\ \hline
\raisebox{0.5em}{$Encrypt$} &\shortstack[l]{$T_{h}+2T_{sm}+T_a+$\\$2T_p+T_{mul}+T_{exp}$} & \shortstack[l]{$T_{h}+2T_{sm}+2T_a+2T_p+$\\$T_{mul}+2T_{exp}+T_{x}+T_{bf}$} & \raisebox{0.5em}{$T_{h}+4T_{sm}+T_a$} & \shortstack[l]{$T_{h}+2T_{sm}+T_a+$\\$2T_p+T_{mul}+T_{exp}$} &\shortstack[l]{$T_{h}+2T_{sm}+T_a+$\\$2T_p+T_{mul}+T_{exp}$}  \\ \hline
$Trapdoor$ & $T_{h}+T_a$ & $T_{h}+T_a$ & $T_{h}+3T_{sm}+T_a$  & $T_{h}+T_a$ & $T_{h}+T_a+T_{sm}$ \\ \hline
\raisebox{0.5em}{$Adjust+Test$} & \shortstack[l]{$2|S|\cdot T_{a}+$\\$T_{mul}+2T_{p}$} & \shortstack[l]{$2|S|\cdot T_{a}+$\\$T_{mul}+2T_{p}$} & \shortstack[l]{$2|S|\cdot T_{a}+$\\$2T_{mul}+4T_p$} & \shortstack[l]{$2|S|\cdot T_{a}+$\\$T_{mul}+2T_{p}$} & \shortstack[l]{$6|S|\cdot T_a + 2|S|\cdot T_{mul} + $\\$4T_{sm} + 2T_{exp} +2T_p$} \\ \hline
\end{tabular}
\caption{The computational cost of KASE: The operation time of hash operations, scalar multiplication, point addition, exclusive or in $\mathbb{G}_T$, exponentiation in $\mathbb{G}_T$, multiplication in $\mathbb{G}_T$, and pairing operation are identified as $T_{h},T_{sm},T_a,T_{x},T_{exp},T_{mul},$ and $T_p$, respectively. Li et al. \cite{li2016verifiable} uses a Bloom filter to verify whether the keyword really exists in the document set. The time taken for the operation is represented as $T_{bf}$. The random generation and integer addition are ignored. The time of $ Adjust + Test $ refers to the computational cost that takes file per one index.}
\label{tab:computation_KASE}
\end{center}
\end{table*}


\begin{table*}[htb]
\begin{center}
\begin{tabular}{c|c|c|c|c|c} \hline 
& Cui et al. \cite{cui2016key} & Li et al. \cite{li2016verifiable} & Zhou et al.\cite{zhou2018file} & First Construction & Main Construction \\ \hline
encrypted keyword & $2|\mathbb{G}|+|\mathbb{G}_T|$ & $2|\mathbb{G}|+3|\mathbb{G}_T|$ & $3|\mathbb{G}|$ & $2|\mathbb{G}|+|\mathbb{G}_T|$ & $3|\mathbb{G}|+2|\mathbb{G}_T|$ \\ \hline
trapdoor & $|\mathbb{G}|$ & $|\mathbb{G}|$ & $2|\mathbb{G}|$ & $|\mathbb{G}|$ & $|\mathbb{G}|+2|\mathbb{Z}_p^*|$ \\ \hline
aggregate key & $|\mathbb{G}|$ & $|\mathbb{G}|$ & $|\mathbb{G}|$ & $|\mathbb{G}|$ & $|\mathbb{G}|$ \\ \hline
\end{tabular}
\caption{The storage cost of KASE: $|\mathbb{G}|, |\mathbb{G}_T|$ refers to the size of $\mathbb{G}, \mathbb{G}_T$.}
\label{tab:cost_KASE}
\end{center}
\end{table*}


The computational cost and the storage size for the first construction are less-than-or-equal to those of the schemes by Cui et al.~\cite{cui2016key} and by Li et al. ~\cite{li2016verifiable} in spite of achieving the provable security, which is an open problem in these works. 

Even in the main construction, the computational cost for the Encrypt algorithm is identical to that of the scheme by Cui et al. ~\cite{cui2016key}, and the computational cost for the Trapdoor algorithm is smaller than that of the scheme by Zhou et al.~\cite{zhou2018file}. Although the computational costs for the Adjust and Test algorithms are greater than those of other schemes, the number of scalar multiplications in $\mathbb{G}$ and exponentiations in $\mathbb{G}_T$, whose computations are heavy, are constant with respect to the number of documents. Thus, the computational cost of the main construction can be considered to be practical. Moreover, for the storage cost, in spite of two additional components in $\mathbb{Z}_p$, the storage size is fairly identical to that of the scheme by Zhou et al. In particular, an element in $\mathbb{G}$ is constructed by two integers on an elliptic curve, i.e., $x$-coordinate and $y$-coordinate, whose bit lengths are the same as the bit length of an element in $\mathbb{Z}_p$. Thus, the entire bit length of $2 \mathbb{Z}_p$ is equal to that of $\mathbb{G}$. The main construction can achieve a similar storage size as other works. 

As will be discussed in detail in the next subsection, the performance of the proposed schemes described above has been achieved as well as the provable security, which is the open problem in the existing works.

\subsection{Security}
In this section, we discuss the security required in KASE. The results are shown in Table~\ref{tab:security}.

\begin{table*}[htb]
\begin{center}
\begin{tabular}{c|c|c|c|c|c} \hline 
& Cui et al. \cite{cui2016key} & Li et al. \cite{li2016verifiable} & Zhou et al.\cite{zhou2018file} & First Construction & Main Construction \\ \hline
Compactness & $\checkmark$ & $\checkmark$ & & $\checkmark$ & $\checkmark$ \\ \hline
Keyword Privacy & & & $\checkmark$ & $\checkmark$ & $\checkmark$ \\ \hline
Aggregate key Unforgeability & & & & $\checkmark$ & $\checkmark$ \\ \hline
Trapdoor Privacy & & &$\checkmark$ & & $\checkmark$ \\ \hline
\end{tabular}
\caption{The Security of KASE: Checkmark means achievement of the provable security.}
\label{tab:security}
\end{center}
\end{table*}

In Table~\ref{tab:security}, the schemes by Cui et al.~\cite{cui2016key} and by Li et al.~\cite{li2016verifiable} do not satisfy the keyword privacy, the aggregate key unforgeability, and the trapdoor privacy. 
On the other hand, the scheme by Zhou et al.~\cite{zhou2018file}, the first construction, and the main construction satisfy the keyword privacy. Although Cui et al. and Li et al. have discussed the keyword privacy informally, their discussions do not include the provable security with reduction algorithms. As a result, the security can be broken through our oracle-based definitions (see the Appendix for details). 

Although the scheme by Zhou et al. satisfies the keyword privacy and the trapdoor privacy, it does not satisfy the compactness. 
Zhou et al.'s scheme assumes a special situation where a remote sensor device encrypts its sensing data. This requires each sensor device to have an extra key for encryption, and the number of keys increases linearly with respect to the number of sensors when viewed across the system. Thus, the compactness cannot be satisfied. We also note that the scheme by Zhou et al. deals with a problem different from ours. Moreover, the aggregate key unforgeability has not been discussed explicitly in other works.


\section{Conclusion} \label{Conclusion}


In this paper, we proposed provably secure KASE scheme and defined the security of KASE formally. To the best of our knowledge, this is the first paper to provide a formal security discussion of KASE. Our main idea was to combine broadcast encryption and aggregate signatures, and we proposed the scheme called first construction by combining the broadcast encryption scheme by Boneh et al. and the aggregate signature scheme by Boneh et al. The security is provably secure with respect to the keyword privacy under the BDHE-assumption and the aggregate key unforgeability under the DHE assumption. Furthermore, by constructing trapdoors that utilize random numbers distributed via secret sharing, we proposed another scheme called main construction that satisfies the trapdoor privacy. We then implemented the proposed schemes and showed that both schemes could encrypt 5,000 keywords within one second. Moreover, a search in the 5,000 keywords can be executed within about three seconds in the first construction and about six seconds in the main construction. These results show that the proposed schemes are practical while achieving provably security. As future work, we plan to propose a generic construction through any broadcast encryption and any aggregate signatures because the proposed schemes in this paper are based on the specific constructions described above. We also plan to optimize implementation, e.g., parallelization of processes, to improve the performance of the proposed schemes.


\appendix

\section{Vulnerability of the Scheme by Cui et al.}

The scheme by Cui et al.~\cite{cui2016key} does not satisfy the keyword privacy because random values embedded in encrypted keywords are common for every keywords. 
We show an example where the scheme by Cui et al. is broken under our definition of the keyword privacy below. 
In the game of the keyword privacy, an adversary can request an encrypted keyword to the encrypt oracle before receiving a challenge encrypted keyword. Then, the adversary chooses a keyword $w_{l'}$ before the guess phase, and requests $c_{i^*,l'}=(c_{1,i^*,l'},c_{2,i^*,l'},c_{3,i^*,l'})$ for encrypt oracle. 
After that, by receiving the challenge encrypted keyword $c_{i^*,\theta}=(c_{1,i^*,\theta},c_{2,i^*,\theta},c_{3,i^*,\theta})$, the adversary can extract the keyword from the challenge encrypted keyword as follows:
\[
c_{3,i^*,\theta}/c_{3,i^*,l'}=e(H(w_\theta),g)/e(H(w_{l'}),g)
\]
Because the adversary knows the challenge keyword and its resulting $H(w_{l^*})$ and $H(w_{l'})$, the adversary can know whether either one of the challenge keyword or the random keyword is encrypted by calculating $e(H(w_{l^*}),g)/e(H(w_{l'}),g)$. 
Thus, when a common random value is embedded in multiple encrypted keywords, i.e., the construction of the scheme by Cui et al., keywords can be extracted from the encrypted keywords. 

Moreover, the scheme by Cui et al. does not satisfy the trapdoor privacy because trapdoor is deterministic with respect to a given keyword. 
We show an example where the scheme by Cui et al. is broken under the definition of the trapdoor privacy. 
In the game of the trapdoor privacy, an adversary can request trapdoors to the trapdoor oracle before receiving a challenge trapdoor. Then, the adversary chooses a keyword $w_{l'}$ before the guess phase, and requests $Tr'=k^*_{agg} \cdot H(w_{l'})$ to the trapdoor oracle. 
After that, by receiving the challenge trapdoor $Tr^*=k^*_{agg} \cdot H(w_\theta)$, the adversary can extract the keyword from the challenge trapdoor as follows: 
\[
(Tr^*\cdot H(w_{l'}))/Tr'=H(w_\theta)
\]
Thus, if trapdoors are deterministic, keywords can be extract from multiple trapdoors.

\section{Proofs of Main Construction}
\label{proof_appendix}
\subsection{Complexity Assumptions}
First, we define the $l$-(D-)BDHE assumption~\cite{boyen2008uber}.
This is an assumption in asymmetric bilinear groups.

\begin{defi}[$(\epsilon, l)$-(D-)BDHE Assumption in $(\mathbb{G}, \mathbb{H}, \mathbb{G}_T)$]
We say the $l$-(D-)BDHE problem in $(\mathbb{G}, \mathbb{H}, \mathbb{G}_T)$ with a security parameter $1^k$ as, 
for a given $(g,g^\alpha,g^{\alpha^2},...,g^{\alpha^l},g^{\alpha^{l+2}},...,\\g^{\alpha^{2l}}, h,h^s$, $\alpha, s\in \mathbb{Z}_p^*$ and $(\mathbb{G}, \mathbb{H}, \mathbb{G}_T)$ as input, 
determining whether $Z\in \mathbb{G}_T$ is $e(g^{\alpha^{l+1}},h^s)$ or a random value $R$. 
We say that a polynomial time algorithm $\mathcal{A}$ can solve the $l$-(D-)BDHE problem in $(\mathbb{G}, \mathbb{H}, \mathbb{G}_T)$ with an advantage $\epsilon$ if the following relation holds:
\begin{eqnarray}
|Pr[\mathcal{A}(g,h,h^s,\bm{y}_{g,h,\alpha,l},e(g_{l+1},h^s))=0]\nonumber \\
-Pr[\mathcal{A}(g,h,h^s,\bm{y}_{g,h,\alpha,l},R)=0]|\ge \epsilon,  \nonumber
\end{eqnarray}
where $\bm{y}_{g,h,\alpha,l}=(g_1,...,g_l,g_{l+2},...,g_{2l})$.
We say the $l$-(D-)BDHE assumption holds in $(\mathbb{G}, \mathbb{H}, \mathbb{G}_T)$ if there is no polynomial-time algorithm that can solve the $l$-(D-)BDHE problem in $(\mathbb{G}, \mathbb{H}, \mathbb{G}_T)$ with $\epsilon$.
\end{defi}

The difference of the assumption described above from the assumption is only the input, i.e., 
$h\in \mathbb{G}$ for the $l$-BDHE assumption and $g^s$ for the $l$-(D-)BDHE assumption. 
Namely, the notation of the input is different.

Next, we define $l$-BDHE assumption and $l$-DHE assumption in asymmetric bilinear groups.
We use these assumptions for the security proofs of the main construction.
We note that the $l$-BDHE assumption in asymmetric bilinear groups is naturally extended from the $l$-(D-)BDHE assumption. 
In particular, in the following assumptions, when $\mathbb{G} = \mathbb{H}$ and $g = h$, the $l$-BDHE assumption in $(\mathbb{G}, \mathbb{H}, \mathbb{G}_T)$ is identical to the $l$-(D-)BDHE assumption in this section and the $l$-DHE assumption in $(\mathbb{G}, \mathbb{H}, \mathbb{G}_T)$ is identical to the $l$-DHE assumption in Section \ref{complexity_assumptoins}.

\begin{defi}[$(\epsilon, l)$-BDHE Assumption in $(\mathbb{G}, \mathbb{H}, \mathbb{G}_T)$]
We say the $l$-BDHE problem in $(\mathbb{G}, \mathbb{H}, \mathbb{G}_T)$ with a security parameter $1^k$ as, 
for a given $(g,g^s,g^\alpha,g^{\alpha^2},...,g^{\alpha^l},g^{\alpha^{l+2}},...,g^{\alpha^{2l}}, \\h,h^s,h^\alpha,h^{\alpha^2},...,h^{\alpha^l},Z)$ with uniformly random $g\in \mathbb{G}, h\in \mathbb{H}$, $\alpha, s\in \mathbb{Z}_p^*$ and $(\mathbb{G}, \mathbb{H}, \mathbb{G}_T)$ as input, 
determining whether $Z\in \mathbb{G}_T$ is $e(g^{\alpha^{l+1}},h^s)$ or a random value $R$. 
We say that a polynomial time algorithm $\mathcal{A}$ can solve the $l$-BDHE problem in $(\mathbb{G}, \mathbb{H}, \mathbb{G}_T)$ with an advantage $\epsilon$ if the following relation holds:
\begin{eqnarray}
|Pr[\mathcal{A}(g,g^s,h,h^s,\bm{y}_{g,h,\alpha,l},e(g_{l+1},h^s))=0]\nonumber \\
-Pr[\mathcal{A}(g,g^s,h,h^s,\bm{y}_{g,h,\alpha,l},R)=0]|\ge \epsilon,  \nonumber
\end{eqnarray}
where $\bm{y}_{g,h,\alpha,l}=(g_1,...,g_l,g_{l+2},...,g_{2l},h_1,...,h_l)$.
We say the $l$-BDHE assumption holds in $(\mathbb{G}, \mathbb{H}, \mathbb{G}_T)$ if there is no polynomial-time algorithm that can solve the $l$-BDHE problem in $(\mathbb{G}, \mathbb{H}, \mathbb{G}_T)$ with $\epsilon$.
\end{defi}

\begin{defi}[$(\epsilon, l)$-DHE Assumption in $(\mathbb{G}, \mathbb{H})$]
We say the $l$-DHE problem in $(\mathbb{G}, \mathbb{H})$ with a security parameter $1^k$ as, 
for a given $(g,g^\alpha,g^{\alpha^2},...,g^{\alpha^l},g^{\alpha^{l+2}},...,g^{\alpha^{2l}},h,h^\alpha,h^{\alpha^2},...,h^{\alpha^l})$ with uniformly random $g\in \mathbb{G}, h\in \mathbb{H}$, $\alpha\in \mathbb{Z}_p^*$ and $(\mathbb{G}, \mathbb{H}, \mathbb{G}_T)$ as input, 
computing $g^{\alpha^{l+1}}$. 
We say that a polynomial time algorithm $\mathcal{A}$ can solve the $l$-DHE problem in $(\mathbb{G}, \mathbb{H})$ with an advantage $\epsilon$ if the following relation holds: 
\[
Pr[\mathcal{A}(g,\bm{y}_{g,h,\alpha,l},g^{\alpha^{l+1}})]\ge \epsilon, 
\]
where $\bm{y}_{g,h,\alpha,l}=(g^\alpha,g^{\alpha^2},...,g^{\alpha^l},g^{\alpha^{l+2}},...,g^{\alpha^{2l}},h^\alpha,h^{\alpha^2},...,\\h^{\alpha^l})$.
We say the $l$-DHE assumption holds in $(\mathbb{G}, \mathbb{H})$ if there is no polynomial-time algorithm that can solve the $l$-DHE problem in $(\mathbb{G}, \mathbb{H})$ with $\epsilon$. 
\end{defi}

\subsection{Proof for Keyword Privacy}

In this section, we show that the main construction satisfies the keyword privacy. 

\begin{theo}[$(\epsilon',n)$-Keyword Privacy]
The main construction satisfies the $(\epsilon',n)$-keyword privacy under the $(\epsilon,n)$-BDHE assumption in $(\mathbb{G}, \mathbb{H}, \mathbb{G}_T)$, where $\epsilon\geq\epsilon'$.
\end{theo}

\textit{Proof}. 
Suppose there exists as adversary $\mathcal{A}$, whose advantage is $\epsilon'$, against the main construction.
We then build an algorithm $\mathcal{B}$ that solves the BDHE problem in $(\mathbb{G}, \mathbb{H}, \mathbb{G}_T)$. 
Let $\mathcal{C}$ be a challenger for the BDHE problem in $(\mathbb{G}, \mathbb{H}, \mathbb{G}_T)$. 
Algorithm $\mathcal{B}$ proceeds as follows.

\begin{itemize}
\item Init:
$\mathcal{A}$ declares a challenge file index $i^* \in [1,n]$ and sends it to $\mathcal{C}$.
\item Setup:
$\mathcal{C}$ sends $(g,g^s,g_1,g_2,...,g_n,g_{n+2},...,g_{2n},h,h^s,\\h_1,h_2,...,h_n,h_{n+2},...,h_{2n},Z)$ to $\mathcal{B}$.
$\mathcal{B}$ randomly generates $sk=\beta$ and calculates $v'=g^\beta g_{i^*}^{-1}$.
$\mathcal{B}$ sends $params=(g,g_1,g_2,...,g_n,g_{n+2},...,g_{2n},h,h_1,h_2,...,h_n)$ to $\mathcal{A}$.
\item Query:
When $\mathcal{A}$ queries for $\mathcal{O}_{Extract}$, $\mathcal{B}$ responds as follows:
\begin{itemize}
\item
If an aggregate key for $i^* \in S$ is queried, returns $\bot$.
\item
If an aggregate key for $i^* \not\in S$ is queried, returns $k_{agg}=(\Pi_{j\in S}g_{n+1-j}^{\beta}) \cdot (\Pi_{j\in S}g_{n+1-j+i^*})^{-1}=\Pi_{j\in S}g_{n+1-j}^{\beta - \alpha^{i^*}}$.
Note that, if $j=i^*$, $(\Pi_{j\in S}g_{n+1-j+i^*})^{-1}$ cannot be calculated, but it can be calculated because of $i^* \not\in S$.
\end{itemize}

When $\mathcal{A}$ queries for $\mathcal{O}_{Encrypt}$, $\mathcal{B}$ randomly generates $t_{i,l}\in \mathbb{Z}_p^*$, calculates the following $c_{i,l}=(c_{1,i,l},c_{2,i,l},c_{3,i,l})$ and responds to $\mathcal{A}$ ($c_{1,i,l}=h^{t_{i,l}},c_{2,i,l}=(v' \cdot g_i)^{t_{i,l}} ,c_{3,i,l}=\frac{e(H(w_{l}),h)^{t_{i,l}}}{e(g_1,h_n)^{t_{i,l}}}$).

\item Guess:
$\mathcal{A}$ declares the challenge keyword $w_{l^*}$ and sends to $\mathcal{B}$.
$\mathcal{B}$ calculates the challenge encrypted keyword $c_{1,i^*,\theta}=h^s,c_{2,i^*,\theta}=(g^s)^\beta ,c_{3,i^*,\theta}=\frac{e(H(w_{l^*}),h^s)}{Z}$.

Then, when $Z=e(g_{n+1},h^s)$, $c_{1,i^*,\theta}=h^s ,c_{2,i^*,\theta}=((g^\beta g_{i^*}^{-1})\cdot g_{i^*})^s=g^{\beta s}=(g^s)^\beta, c_{3,i^*,\theta}=\frac{e(H(w_{l^*}),h)^s}{e(g_1,h_n)^s}=\frac{e(H(w_{l^*}),h^s)}{e(g_{n+1},h^s)}=\frac{e(H(w_{l^*}),h^s)}{Z}$.
Therefore, the calculation results are identical to the Encrypt algorithm of the main construction. 
$\mathcal{B}$ sends $c_{i^*,\theta}=(c_{1,i^*,\theta},c_{2,i^*,\theta},c_{3,i^*,\theta})$ to $\mathcal{A}$.
$\mathcal{A}$ chooses $\theta'\in \{0,1\}$ and sends it to $\mathcal{B}$.
Then, $\mathcal{B}$ sends $\theta'$ to $\mathcal{C}$ as a guess of $\theta$.
\end{itemize}
In the guess phase, if $Z$ is a random value, then $\Pr[\theta=\theta']=1/2$. 
On the other hand, if $Z=e(g_{n+1},h^s)$, $|\Pr[\theta=\theta']-1/2|>\epsilon'$. 
This indicates that $\mathcal{B}$ has an advantage over $\epsilon'$ for solving the $(\epsilon,n)$-BDHE problem in $(\mathbb{G}, \mathbb{H}, \mathbb{G}_T)$. 
Thus, if the $(\epsilon,n)$-BDHE assumption holds in $(\mathbb{G}, \mathbb{H}, \mathbb{G}_T)$, the main construction satisfies the $(\epsilon',n)$-keyword privacy.

\subsection{Proof for Aggregate key Unforgeability}
In this section, we show that the main construction satisfies the aggregate key unforgeability. 

\begin{theo}[$(\epsilon',n)$-Aggregate key Unforgeability]
The main construction satisfies the $(\epsilon',n)$-aggregate key unforgeability under the $(\epsilon,n)$-DHE Assumption in $(\mathbb{G}, \mathbb{H})$, where $\epsilon=\epsilon'$.
\end{theo}

$Proof.$
Suppose there exists as adversary $\mathcal{A}$, whose advantage is $\epsilon'$, against the main construction.
We then build an algorithm $\mathcal{B}$ that solves the DHE problem. 
Let $\mathcal{C}$ be a challenger for the DHE problem. 
Algorithm $\mathcal{B}$ proceeds as follows.

\begin{itemize}
\item Setup:
$\mathcal{C}$ sends $(g,g_1,g_2,...,g_n,g_{n+2},...,g_{2n},h,h_1,h_2,\\...,h_n,h_{n+2},...,h_{2n})$ to $\mathcal{B}$.
$\mathcal{B}$ randomly generates $sk=\beta$ and calculates $v'=g^\beta g_{i^*}^{-1}$.
$\mathcal{B}$ sends $params=(g,g_1,g_2,...,g_n,g_{n+2},...,g_{2n},h,h_1,h_2,...,h_n)$ to $\mathcal{A}$.
\item Query:
When $\mathcal{A}$ queries for $\mathcal{O}_{Extract}$, $\mathcal{B}$ responds as follows:
\begin{itemize}
\item
If an aggregate key for $i^* \in S$ is queried, returns $\bot$.
\item
If an aggregate key for $i^* \not\in S$ is queried, returns $k_{agg}=(\Pi_{j\in S}g_{n+1-j}^{\beta}) \cdot (\Pi_{j\in S}g_{n+1-j+i^*})^{-1}=\Pi_{j\in S}g_{n+1-j}^{\beta - \alpha^{i^*}}$.
Here, if $j=i^*$, then $(\Pi_{j\in S}g_{n+1-j+i^*})^{-1}$ cannot be calculated, but it can be calculated because of $i^* \not\in S$.
\end{itemize}

When $\mathcal{A}$ queries for $\mathcal{O}_{Encrypt}$, $\mathcal{B}$ randomly generates $t_{i,l}\in \mathbb{Z}_p^*$, calculates $c_{i,l}=(c_{1,i,l},c_{2,i,l},c_{3,i,l})$ and responds to $\mathcal{A}$ ($c_{1,i,l}=h^{t_{i,l}},c_{2,i,l}=(v' \cdot g_i)^{t_{i,l}} ,c_{3,i,l}=\frac{e(H(w_{l}),h)^{t_{i,l}}}{e(g_1,h_n)^{t_{i,l}}}$).

\item Forge:
$\mathcal{A}$ outputs $S^*,k_{agg}^*$ and sends them to $\mathcal{B}$.
\begin{itemize}
\item If $i^*\not\in S^*$, abort
\item If $i^*\in S^*$, $k_{agg}^*=(\Pi_{j\in S,j\not= i^*}g_{n+1-j})^{\beta -\alpha^{i^*}}\cdot (g_{n+1-i^*})^{\beta- \alpha^{i^*}}$.
By using this $k_{agg}^*$, $\mathcal{B}$ calculates $(\Pi_{j\in S^*,j\not= i^*}g_{n+1-j})^\beta \cdot (\Pi_{j\in S^* ,j\not= i^*}g_{n+1-j+i^*})^{-1} \cdot (g_{n+1-i^*})^\beta/k_{agg}^*=g^{\alpha+1}$ and outputs results.
\end{itemize}
\end{itemize}
The result in the above $(\epsilon',n)$-aggregate key unforgeability game is identical to the answer of the $(\epsilon,n)$-DHE problem in $(\mathbb{G}, \mathbb{H})$. 
That is, the advantage of the aggregate key unforgeability game is equal to the advantage of the $(\epsilon,n)$-DHE problem in $(\mathbb{G}, \mathbb{H})$. 
Thus, if the $(\epsilon,n)$-DHE assumption holds, the main construction satisfies the $(\epsilon',n)$-aggregate key unforgeability.

\begin{IEEEbiography}[{\includegraphics[width=1in,height=1.25in,clip,keepaspectratio]{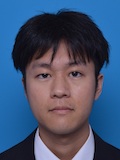}}]{Masahiro Kamimura} received B.Eng degree in Engineering Science from Osaka University, Japan, in 2018. He has recently joined M.S. course in Graduate School of Information Science and Technology in Osaka University, Japan. His research interests include information security.
\end{IEEEbiography}

\begin{IEEEbiography}[{\includegraphics[width=1in,height=1.25in,clip,keepaspectratio]{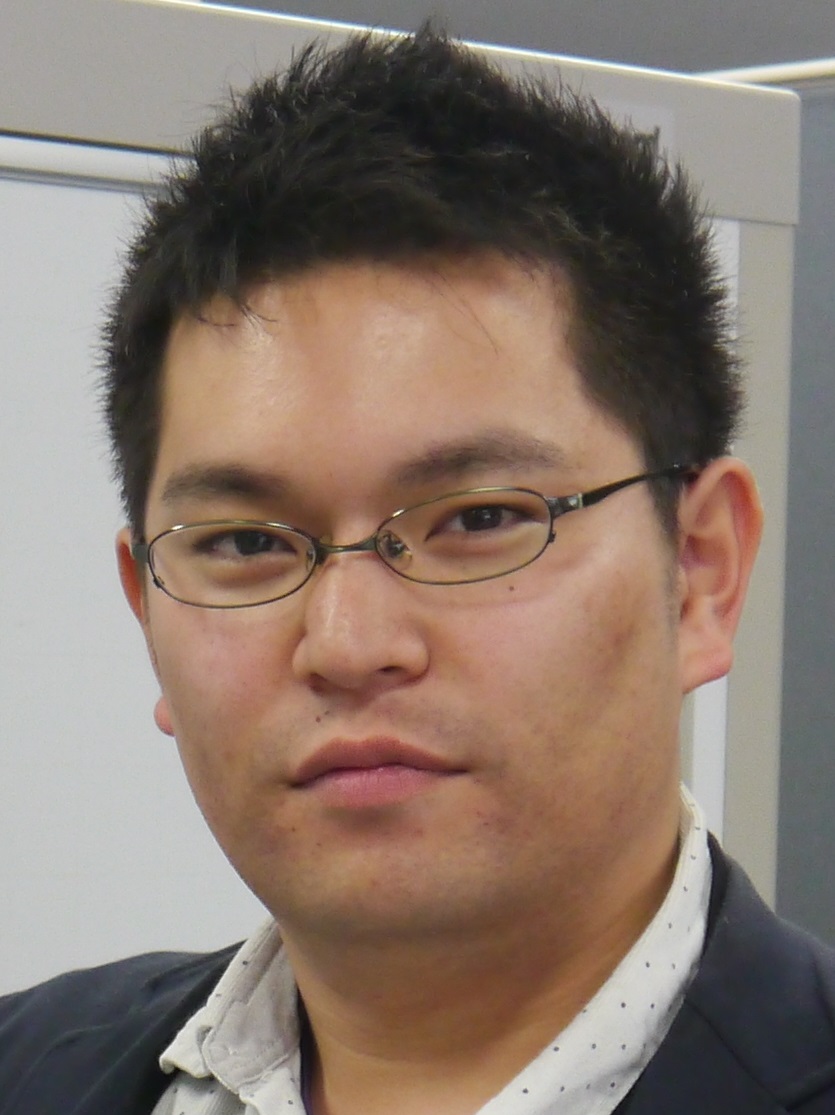}}]{Naoto Yanai} 
received the B.Eng. degree from The National Institution of Academic Degrees and University Evaluation, Japan, in 2009, the M.S.Eng. from the Graduate School of Systems and Information Engineering, the University of Tsukuba, Japan, in 2011, and the Dr.E. degree from the Graduate School of Systems and Information Engineering, the University of Tsukuba, Japan, in 2014. He is an assistant professor at Osaka University, Japan. His research area is cryptography and information security.
\end{IEEEbiography}

\begin{IEEEbiography}[{\includegraphics[width=1in,height=1.25in,clip,keepaspectratio]{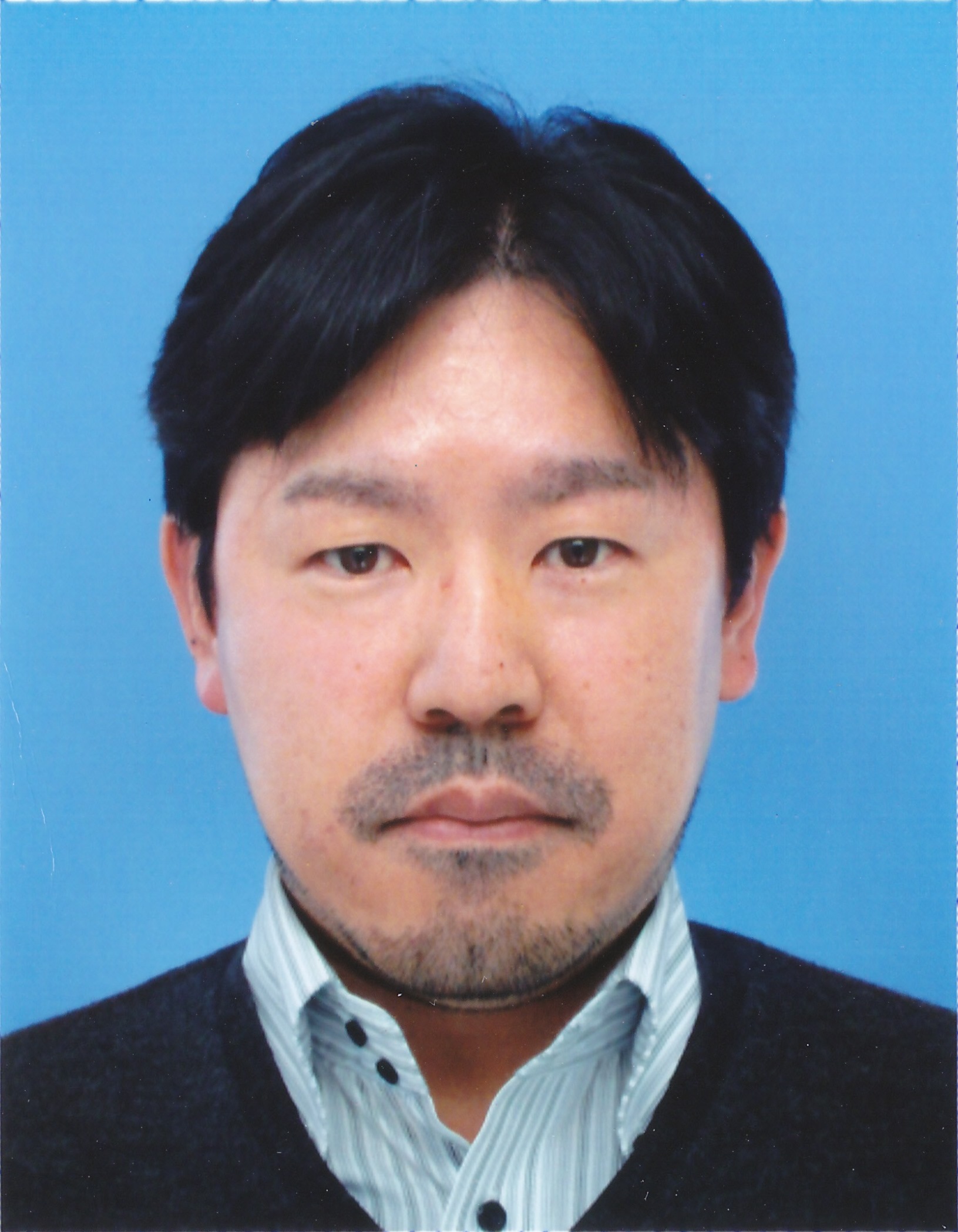}}]{Shingo Okamura} received his B.E., M.E., and Ph.D. degrees in information science and technology from Osaka University in 2000, 2002, and 2005, respectively. Since 2005, he has worked for Osaka University. In 2008, he joined National Institute of Technology, Nara College. Currently, he is an associate professor at the college. His research interests include cryptographic protocols and cyber security. He is a member of IEICE, IEEJ, ACM, IEEE, and IACR. \end{IEEEbiography}

\begin{IEEEbiography}[{\includegraphics[width=1in,height=1.25in,clip,keepaspectratio]{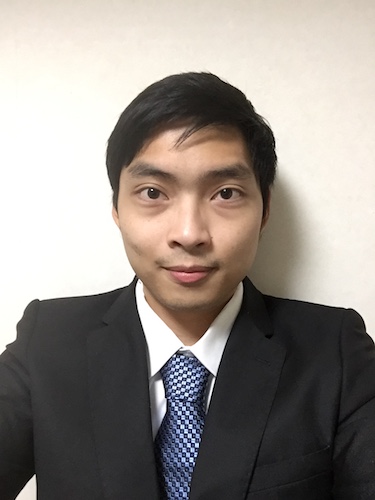}}]{Jason Paul Cruz} received his B.S. degree in Electronics and Communications Engineering and M.S. degree in Electronics Engineering from the Ateneo de Manila University, Quezon City, Philippines, in 2009 and 2011, respectively, and his Ph.D. degree in Engineering from the Graduate School of Information Science, Nara Institute of Science and Technology, Nara, Japan in 2017. He is currently a Specially Appointed Assistant Professor at Osaka University, Osaka, Japan. His current research interests include role-based access control, blockchain technology, hash functions and algorithms, privacy-preserving cryptography, and Android programming.
\end{IEEEbiography}

\EOD

\end{document}